\documentclass[journal]{IEEEtran}
\usepackage{cite}
\usepackage{amsmath,amssymb,amsfonts}
\usepackage{algorithmic}
\usepackage{graphicx}
\usepackage{textcomp}
\usepackage{xcolor}
\usepackage{subfigure}
\usepackage{graphicx}
\usepackage{caption2}

\newtheorem{lemma}{Lemma}
\newtheorem{fact}{Fact}

\newtheorem{theorem}{Theorem}

\ifCLASSINFOpdf
\else
\fi
\begin{document}
%
\title{Design of Artificial Interference Signals for Covert Communication Aided by Multiple Friendly Nodes}
%
%
%

\author{Xuyang~Zhao,
        Wei~Guo,
        and~Yongchao~Wang,
\thanks{}}

%
%

\markboth{ }%
{Shell \MakeLowercase{\textit{et al.}}: Bare Demo of IEEEtran.cls for IEEE Journals}
%



\maketitle

\begin{abstract}
In this paper, we consider a scenario of covert communication aided by multiple friendly interference nodes. The objective is to conceal the legitimate communication link under the surveillance of a warden. The main content is as follows: first, we propose a novel strategy for generating artificial noise signals in the considered covert scenario. Then, we leverage the statistical information of channel coefficients to optimize the basis matrix of the artificial noise signals space in the absence of accurate channel fading information between the friendly interference nodes and the legitimate receiver. The optimization problem aims to design artificial noise signals within the space to facilitate covert communication while minimizing the impact on the performance of legitimate communication. Second, a customized Rimannian Stochastic Variance Reduced Gradient (R-SVRG) algorithm is proposed to solve the non-convex problem. In the algorithm, we employ the Riemannian optimization framework to analyze the geometric structure of the basis matrix constraints and transform the original non-convex optimization problem into an unconstrained problem on the complex Stiefel manifold for solution. Third, we theoretically prove the convergence of the proposed algorithm to a stationary point. In the end, we evaluate the performance of the proposed strategy for generating artificial noise signals through numerical simulations. The results demonstrate that our approach significantly outperforms the Gaussian artificial noise strategy without optimization.
\end{abstract}


%
\IEEEpeerreviewmaketitle

\section{Introduction}

With the development of wireless communication technology, modern society increasingly relies on information transmission. At the same time, the open nature of wireless channels makes information vulnerable to interception and attacks by unauthorized users, leading to security threats during transmission. In recent years, various security incidents caused by wireless communication have occurred repeatedly. In the military field, once electromagnetic signals are intercepted by enemy unauthorized users, the subsequent decryption of our information or attacks on communication links by the enemy will have serious consequences. Therefore, issues related to wireless communication security cannot be ignored \cite{wireless-security-survey}.

The traditional solution for wireless communication security is based on cryptographic encryption techniques. When meaningful information is encrypted, it becomes random and loses its original meaning. This failure to conceal ongoing communication activities attracts the attention of unauthorized users, posing a risk of information leakage. Instead, a strategy is to embed secret messages into public carrier objects for communication using steganography \cite{Steganography}. Drawing inspiration from steganography, communications parties need to seek communication strategies with low detection probabilities.

Recently, covert communication has attracted widespread attention due to its ability to conceal communication behaviors \cite{covert-fundamental-limits}. In covert communication, the transmitter (Alice) attempts to reliably communicate with the legitimate receiver (Bob), without being detected by the warden (Willie), who is observing the communication environment. In \cite{square-root-law}, a square-root law for covert communication in additive white Gaussian noise (AWGN) channels was established for the first time. This law stipulates that under the surveillance of a warden, at most $\mathcal{O}(\sqrt{n})$ information bits can be reliably and covertly transmitted to the intended receiver in $n$ channel uses. Furthermore, the square root law (SRL) has been extended to binary symmetric channels (BSC) \cite{SRL-BSC}, discrete memoryless channels (DMC) \cite{DMC-SRL}-\cite{DMC-SRL-Zheng}, multiple access channels \cite{MAC-SRL}, broadcast channels \cite{Broadcost-SRL}, and multiple-input multiple-output (MIMO) AWGN channels \cite{MIMO-AWGN-SRL}. However, when $n \rightarrow \infty$, the achievable covert transmission rate approaches zero. The square root law provides the theoretical limit of communication under covert transmission behavior undetectable by the warden, but this covert channel represents a zero-rate channel.

Research has shown that it is possible to achieve positive covert transmission rates when Willie faces uncertainties such as transmission time \cite{transmission-time}, channel noise power \cite{channel-noise-power}, communication environment interference \cite{environment_interference}, Alice's transmission time, and interference time \cite{continuous-time}. In existing works \cite{noise-uncertainty}, covert communication under bounded uncertainty noise interference and unbounded uncertainty noise interference has been studied, deriving the maximum transmission rate given covert requirements. \cite{poisson-interference} investigated covert communication in interference Poisson fields, analyzing the impact of concurrent interference source density and transmission power on covert throughput under requirements of covert and reliability. \cite{MIMO-Poisson} investigates the covert communication performance of centralized and distributed multiple-antenna transmitters under the covert interference in Poisson point process (PPP) networks, where multiple wardens are monitoring.

However, wireless transmission inevitably suffers from co-channel interference, where interference in the channel not only reduces the detection performance of Willie but also affects normal wireless communication. The application of artificial noise techniques can mitigate the impact of co-channel interference on the communication performance between legitimate transceivers. In \cite{non-cooperative}, a non-cooperative interference machine selection scheme was proposed in scenarios where multiple friendly nodes assist covert communication. \cite{cognitive-jammer} employed cognitive jammer to sense the transmission of Alice, dynamically adjusting interference strategies to achieve covert performance gains. In \cite{multi-antenna-jammer}, the interference emitter equipped with multiple antennas is studied. When Willie's channel state information (CSI) is known, the optimal strategy for the interference emitter is to use all available power to transmit interference signals toward Willie. When Willie's CSI is unknown, the optimal strategy for the interference emitter is to send interference signals to Bob's null space. In the case of full-duplex receivers, \cite{full-duplex-jammer} assumes that the warden possesses perfect channel information. It considers the prior probabilities transmitted by the sender and the range of values for receiver interference power as optimization variables. It maximizes the warden's probability of error detection under a given effective transmission rate requirement, revealing the enhancement in covert performance by leveraging prior probabilities as controllable factors.

For existing artificial noise (AN) strategies, when only perfect channel state information (CSI) of the legitimate link is available, artificial noise located in the null space of the legitimate channel is selected to mitigate the impact of artificial noise on legitimate communication performance \cite{secure-MIMO}-\cite{AN-cooperative-relay}. Existing works on artificial noise strategy design in covert communication require perfect CSI of the channel between the jammers and the legitimate receiver. To the best of the author's knowledge, currently there is no research investigating the design of artificial noise signals in covert communication when perfect CSI is unknown.

In this paper, we propose for the first time a novel strategy for generating artificial noise signals in the scenario of covert communication aided by multiple friendly nodes, without perfect knowledge of the CSI. The strategy aims to significantly degrade Willie's detection performance while minimizing interference to Bob. Additionally, a customized R-SVRG algorithm is developed for solving the problem. Its main contributions are as follows:

\begin{itemize}
\item {\it Artificial noise model construction and expected model reformulation:} we model the artificial noise signals in the scenario of covert communication aided by multiple friendly interference nodes as random vectors in a specific subspace within the entire space. By designing the basis matrix in this subspace, we aim to find the artificial noise signal space that can significantly degrade Willie's detection performance while minimizing the impact on Bob. Based on imperfect CSI, we construct an expected model and reformulate it into minimizing the average of a large, but finite, number of loss functions under unitary matrix constraints.
\item {\it Riemannian manifold reformulation framework:} The non-convex optimization problem constructed is addressed using the Riemannian manifold optimization framework different from the traditional Euclidean optimization framework. We analyze the geometric structure of constraints in the optimization problem and embed the constraints into the search space, transforming them into unconstrained optimization problems on the complex Stiefel manifolds for resolution.
\item {\it Stochastic gradient solving algorithm:} We customize the R-SVRG algorithm on the complex Stiefel manifold to solve the problem of minimizing the average of the large, but finite, number of loss functions on the manifold considered in this paper. By leveraging the geometric structure of the constraints, this algorithm offers a more effective solution to the optimization problem and has the advantage of lower computational burden.
\item {\it Theoretically-guaranteed performance:} We provide certain parameter selection strategies and theoretical convergence analysis for the proposed algorithm, demonstrating that the algorithm can converge to stationary points. Finally, we conduct simulations to verify the convergence of the algorithm and the enhancement of covert performance by our proposed artificial noise strategy.
\end{itemize}

The rest of the paper is organized as follows: In Section II, we describes the covert communication scenario aided by multiple friendly nodes considered in this paper, along with the channel model and artificial noise signal model in the scenario. Furthermore, based on binary hypothesis testing theory, the average minimum detection error probability at Willie is analyzed as a quantification metric for covert performance, leading to the formulation of the artificial noise matrix design problem. In Section III, the structure of constraints is analyzed, and the constraints of the optimization problem are embedded into the search space to construct the corresponding Riemannian manifold. Additionally, a customized R-SVRG algorithm is employed to solve the optimization problem. The theoretical analysis of the convergence of the proposed algorithm is presented in Section IV. Simulation results are presented in Section V. Finally, conclusions are drawn in Section VI.

{\it Notations:} Scalar variables are represented by standard lowercase letters; vectors and matrices are denoted by lowercase and uppercase bold symbols, respectively; $\mathbb{R}$ and $\mathbb{C}$ denote the sets of real and complex numbers, respectively; $\mathbf{a}[t]$ denotes the $t$-th sample realization obtained by sampling from the random vector $\mathbf{a}$; $(\cdot)^T$, $(\cdot)^H$, $\operatorname{Tr}(\cdot)$, $\mathbb{E}(\cdot)$ represent transpose, conjugate transpose, matrix trace, and expectation, respectively; $|\cdot|$ denotes the absolute value and $\|\cdot\|_2$ denotes the Euclidean vector norm; $\mathbf{I}_m$ denotes the $m \times m$ identity matrix; $\nabla \left(\cdot\right)$ represents the Euclidean gradient of a function; $\operatorname{grad}\left(\cdot\right)$ represents the Riemannian gradient of a function; $\operatorname{Re}\left(\cdot\right)$ takes the real part of the complex variable; $\operatorname{Im}\left(\cdot\right)$ takes the image part of the complex variable; $\mathcal{CN}(0, \sigma^2)$ denotes the circularly symmetric complex Gaussian distribution with zero mean and variance $\sigma^2$.

\section{SYSTEM MODEL AND PROBLEM FORMULATION}

\subsection{Communication Scenario and Assumptions}

\begin{figure}[!t]
\centering
\includegraphics[width=3.5in]{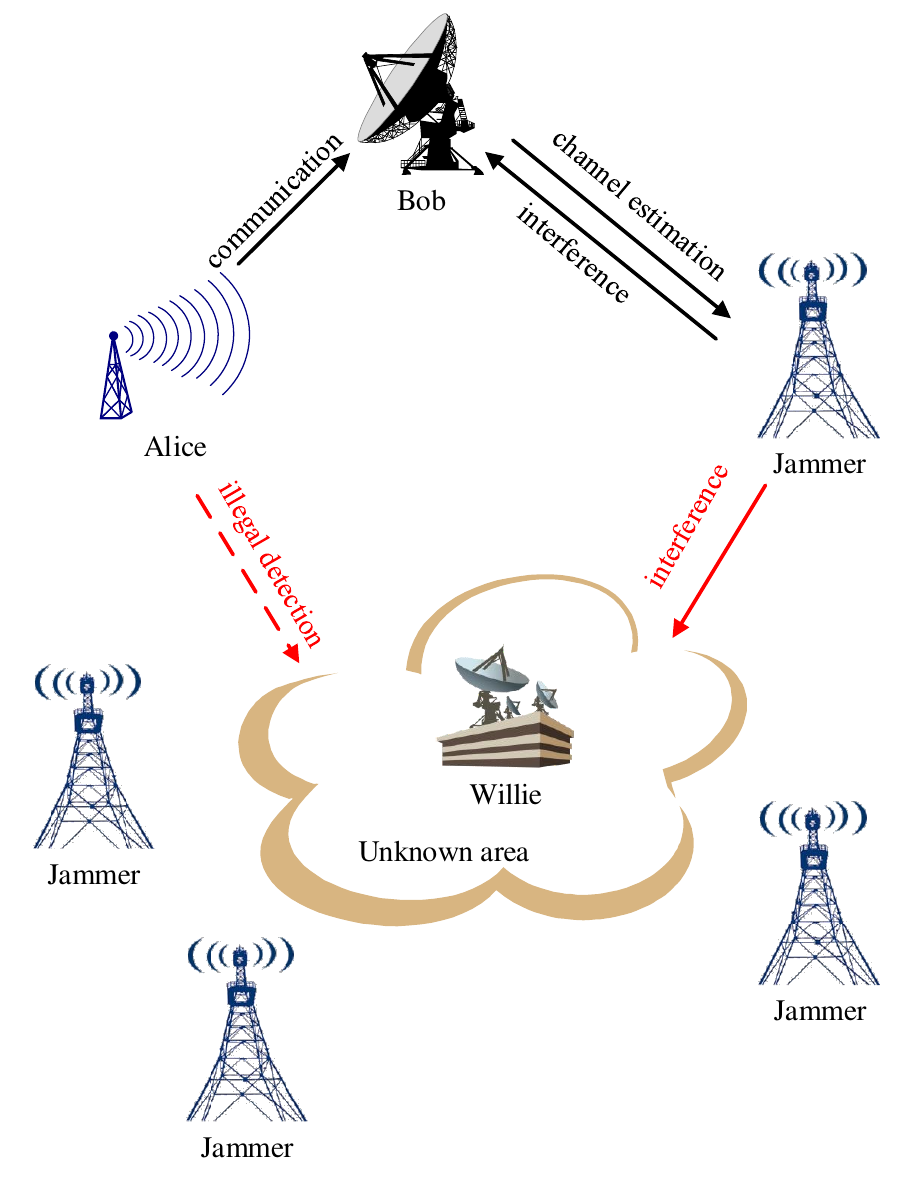}
\caption{Illustration of the covert communication scenario assisted by multiple friendly interference nodes.}
\label{covert scenario}
\end{figure}

As shown in Fig.~\ref{covert scenario}, we consider a multiple friendly interference nodes assisted covert communication system, where the transmitter (Alice) would like to send the information to the legitimate user (Bob) with the assistance of multiple friendly interference nodes without being detected by the warden (Willie). Additionally, for generality, it is assumed that there exist $n(n>1)$ friendly interference nodes acting as jammers in the space, transmitting artificial noise to assist Alice and Bob in covert communication. Suppose that each of Alice, Bob, jammers and Willie is equipped with a single antenna. When Alice communicates with Bob, she transmits her message by mapping it to the complex sequence $s_a\sim\mathcal{C N}\left(0,1\right)$, while $n$ jammers transmit complex sequences $\boldsymbol{v}_j=\left[v_1,v_2,...,v_n\right]^T$ to assist in covert communication between Alice and Bob. Here, $v_k$ represents the AN signals transmitted by the $k$-th jammer. For simplicity, let the average power per symbol in $s$ and $\boldsymbol{v}_j$ is normalized to $1$. Alice transmits with power $P_a$, and each jammer transmits with power $P_j$.

In this work, we define the $h_{a,b}$, $h_{j_k,b}$, $h_{a,w}$, $h_{j_k,w}\in \mathbb{C}$ are the channel coefficients for Alice-Bob, $k$-th jammer-Bob, Alice-Willie and $k$-th jammer-Willie. We consider quasi-static fading channels, hence the channel coefficients remain unchanged in one block and change independently from one block to another.

We assume that the accurate channel coefficient $h_{a,b}$ between Alice and Bob is known while there exists estimation error in the estimated channel coefficient between the jammers and Bob. Here $h_{a,b}$ is considered a known complex constant. Considering the uncertainty of the channel coefficients, the channel coefficient between the $k$-th jammer and Bob is given by \cite{fading-channel-uncertain}

\begin{equation}\label{channel model}
h_{j_k, b}=\widehat{h_{j_k, b}}+\widetilde{h_{j_k, b}}
\end{equation}
where $\widehat{h_{j_k, b}}$ is the known part of $h_{j_k, b}$, and $\widetilde{h_{j_k, b}}\sim \mathcal{C N}\left(0,\sigma_{j_k,b}^2\right)$ is the uncertain part of $h_{j_k, b}$.

For the passive warden Willie, legitimate nodes can hardly obtain accurate information about the channel coefficients related to Willie. However, based on experience, they acquire statistical information about the channel coefficients between Alice and Willie, as well as between each jammer and Willie, denoted as $h_{a,w}\sim\mathcal{C N}\left( 0,\sigma_{a,w}^2 \right)$, $h_{j_k,w}\sim\mathcal{C N}\left( 0,\sigma_{j_k,w}^2 \right)$.

\subsection{Artificial Noise Signal Model}

In the scenario considered in this paper, the traditional Gaussian artificial noise signal can be described as follows: randomly select a vector $\mathbf{c}=\left[c_1,c_2,...,c_n\right]^T$ in the complex vector space $\mathbb{C}^n$, where the $k$-th element $c_k$ serves as the AN signal sent by the $k$-th jammer in once channel use. Mathematically, this is represented as $\mathbf{c}=\mathbf{E}\mathbf{b}$, where $\mathbf{E} = \left[\mathbf{e}_1,\mathbf{e}_2,...,\mathbf{e}_n\right]$, $\mathbf{b} \sim \mathcal{C N}\left( 0,\mathbf{I}_n \right)$.

This work aims to design artificial noise random signals sent by $n$ jammers, minimizing interference power to Bob while deteriorating Willie's detection performance. Specifically, the goal is to design a space of AN signals that meet the above objectives, enabling random selection of an artificial noise signal from this space to assist covert communication between Alice and Bob. Mathematically, this space can be regarded as a subspace of the complex vector space $\mathbb{C}^n$, denoted as $\operatorname{Span}(\mathbf{x}_1,\mathbf{x}_2,...,\mathbf{x}_p)$, where $\mathbf{X}=\left[\mathbf{x}_1,\mathbf{x}_2,...,\mathbf{x}_p\right] \in \mathbb{C}^{n \times p}$ represents the basis matrix of the space of AN signals sent by $n$ jammers, and $\mathbf{v}_j \in \operatorname{Span}(\mathbf{x}_1,\mathbf{x}_2,...,\mathbf{x}_p)$ represents the AN noise sent by $n$ jammers. The artificial noise $\mathbf{v}_j$ can be modeled as

\begin{equation}\label{AN signal model}
\mathbf{v}_j=\left[v_1,v_2,...,v_n\right]=\mathbf{Xa}
\end{equation}
where the element $v_k$ represent the AN signal sent by the $k$-th jammer in once channel use, $\mathbf{a} \sim \mathcal{C N}\left(0,\mathbf{I}_p\right)$ denotes randomly generated coefficients for linear combinations, and $\mathbf{X}^H\mathbf{X}=\frac{n}{p}\mathbf{I}_p$. Consequently, the original design of AN signals is transformed into the design of the basis matrix for the AN space.

Remark: According to the average power per symbol in $\boldsymbol{v}_j$ is normalized to $1$, and $\mathbf{X}$ being a basis matrix, i.e. the matrix $\mathbf{X}$ has full column rank, we let $\mathbf{a} \sim \mathcal{C N}\left(0,\mathbf{I}_p\right)$, and $\mathbf{X}^H\mathbf{X}=\frac{n}{p}\mathbf{I}_p$ can be easily obtained from
\begin{equation}\label{per symbol power}
\mathbb{E}\left[\left|v_k\right|^2\right]=\frac{1}{n} \mathbb{E}\left[\|\mathbf{X} \mathbf{a}\|_2^2\right]=\frac{1}{n} \operatorname{Tr}\left(\mathbf{X}^H \mathbf{X} I_p\right)=1
 \end{equation}

\subsection{Covert Performance}
In covert communication scenario, Alice's objective is to communicate with Bob without being detected by the warden Willie. Willie receives electromagnetic signals from the space and determines whether communication is taking place between Alice and Bob based on the observation results. Mathematically, Willie's detection process can be described as Willie distinguishing between two hypotheses by applying a specific decision rule. One hypothesis, denotes as $\mathcal{H}_1$, assumes that Alice is sending covert messages to Bob. The alternative hypothesis, denoted as $\mathcal{H}_0$, assumes that Alice is not sending covert messages to Bob. Therefore, the signal received by Willie can be written as

\begin{equation}\label{Willie signals}
y_w=\left\{\begin{array}{cc}
\sqrt{P_j} \mathbf{h}_{j, w}^H \mathbf{v}_j+n_w & \mathcal{H}_0  \\
\sqrt{P_a} h_{a, w} s_a+\sqrt{P_j} \mathbf{h}_{j, w}^H \mathbf{v}_j+n_w & \mathcal{H}_1
\end{array}\right.
\end{equation}
where $\mathbf{h}_{j,w}=\left[h_{j_1,w},h_{j_2,w},...,h_{j_n,w}\right]^H$, and $n_w \sim \mathcal{C N}\left(0,\sigma_w^2\right)$ is the received noise at Willie.

The variance of $y_w$, which is referred to as $\lambda=\mathbb{E}\left[\left(y_w-\mathbb{E}\left(y_w\right)\right)^2\right]$, under different hypotheses are

\begin{subequations}
\begin{align}
\mathcal{H}_0:& \lambda_0=P_j \mathbf{h}_{j, w}^H \mathbf{X} \mathbf{X}^H \mathbf{h}_{j, w}+\sigma_w^2 \label{H0 variance} \\
\mathcal{H}_1:& \lambda_1=P_a\left|h_{a, w}\right|^2+P_j \mathbf{h}_{j, w}^H \mathbf{X} \mathbf{X}^H \mathbf{h}_{j, w}+\sigma_w^2 \label{H1 variance}
\end{align}
\end{subequations}

{\it Proof:} Please see Appendix~\ref{lambda proof}.

Willie determines whether there is covert communication occurring between Alice and Bob by distinguishing between hypotheses $\mathcal{H}_1$ and $\mathcal{H}_0$. We use $\mathbb{P}_{FA}$ and $\mathbb{P}_{MD}$ to respectively represent the probabilities of false alarm (FA) and missed detection (MD), and utilize Willie's detection error probability to assess the covert performance. We express the detection error probability $\mathbb{P}_e$ as
\begin{equation}\label{detection error probability}
\mathbb{P}_e=\mathbb{P}_{FA}+\mathbb{P}_{MD}
\end{equation}

We assume Willie applies the maximum likelihood detection. According to the theorem in \cite{test-hypotheses}, the optimal detector has

\begin{equation}\label{P-FA+P-MD}
\mathbb{P}_{FA}+\mathbb{P}_{MD} = 1-\mathcal{V}_T\left(\mathbb{P}_1,\mathbb{P}_0\right)
\end{equation}
where $\mathbb{P}_0$ is the probability distribution of $y_w$ when $\mathcal{H}_0$ is true, $\mathbb{P}_1$ is the probability distribution of $y_w$ when $\mathcal{H}_1$ is true and $\mathcal{V}_T\left(\mathbb{P}_0,\mathbb{P}_1\right)$ is the total variation distance between $\mathbb{P}_0$ and $\mathbb{P}_1$.

According to Pinsker's inequality \cite{square-root-law}, we can get the lower bound of detection error probability
\begin{equation}\label{lower bound of Pe}
\mathbb{P}_{FA}+\mathbb{P}_{MD} = 1-\mathcal{V}_T\left(\mathbb{P}_1,\mathbb{P}_0\right) \geq 1-\sqrt{\frac{1}{2} \mathcal{D}\left(\mathbb{P}_1 \| \mathbb{P}_0\right)}
\end{equation}
where $ \mathcal{D}\left(\mathbb{P}_1 \| \mathbb{P}_0\right)$ denotes the KL divergence between the probability densities of $y_w$ under $\mathcal{H}_1$ and $\mathcal{H}_0$.

The KL divergence is defined as \cite{information-theory}

\begin{equation}\label{KL}
\mathcal{D}\left(\mathbb{P}_0 \| \mathbb{P}_1\right) = \int_\chi p_0(x) \ln \frac{p_0(x)}{p_1(x)} d x
\end{equation}
where $ p_0(x)$ and $p_1(x)$ are the densities of $\mathbb{P}_1$ and $\mathbb{P}_0$, and $\chi$ is the support of $p_1(x)$.

Furthermore, $ \mathcal{D}\left(\mathbb{P}_1 \| \mathbb{P}_0\right)$ are respectively given as \cite{robust-beamforming}

\begin{equation}\label{KL divergence}
\begin{aligned}
\mathcal{D}\left(\mathbb{P}_1 \| \mathbb{P}_0\right) & =\ln \frac{\lambda_1}{\lambda_0}+\frac{\lambda_0}{\lambda_1}-1 \\
& =\ln \left(P_a\left|h_{a, w}\right|^2\!+\!P_j \mathbf{h}_{j, w}^H \mathbf{X} \mathbf{X}^H \mathbf{h}_{j, w}\!+\!\sigma_w^2\right)\\
&\ \ \ -\ln \left(P_j \mathbf{h}_{j, w}^H \mathbf{X} \mathbf{X}^H \mathbf{h}_{j, w}+\sigma_w^2\right)\\
&\ \ \ -\frac{P_a\left|h_{a, w}\right|^2}{P_a\left|h_{a, w}\right|^2 + P_j \mathbf{h}_{j, w}^H \mathbf{X} \mathbf{X}^H \mathbf{h}_{j, w}+\sigma_w^2}
\end{aligned}
\end{equation}

According to \eqref{lower bound of Pe}, it is easy to see that the smaller the value of $ \mathcal{D}\left(\mathbb{P}_1 \| \mathbb{P}_0\right)$, the closer $\mathbb{P}_{FA}+\mathbb{P}_{MD}$ gets to $1$, indicating better covert performance. Since both $h_{a,w}$ and $\mathbf{h}_{j,w}$ are random variables, we choose $\mathbb{E}\left[\mathcal{D}\left(\mathbb{P}_1 \| \mathbb{P}_0\right)\right]$ as the measure of covert performance in this paper.

\subsection{Optimization Problem Formulation}

When covert communication occurs, the signal received by Bob is represented as
\begin{equation}\label{Bob signals}
y_b = \sqrt{P_a}h_{a,b}s_a+\sqrt{P_j}\mathbf{h}_{j,b}^Hv_j+n_b
\end{equation}
where $\mathbf{h}_{j,b}=\left[h_{j_1,b},h_{j_2,b},...,h_{j_n,b}\right]^H$, and $n_b \sim \mathcal{C N}\left(0,\sigma_b^2\right)$ is the received noise at Bob.

Then, the covert rate at Bob can be expressed as
\begin{equation}\label{Bob covert rate}
R_b=\frac{1}{2} \log \left(1+\frac{P_a \left| h_{a,b}\right|^2}{P_j\mathbb{E}\left(\mathbf{h}_{j,b}^H\mathbf{X}\mathbf{X}^H\mathbf{h}_{j,b}\right)+\sigma_b^2}\right)
\end{equation}

In the scenario considered in this paper, we aim to maximize $R_b$ while maximization $\mathbb{P}_{FA}+\mathbb{P}_{MD}$. According \eqref{KL divergence} and \eqref{Bob covert rate}, the covert rate maximization problem is formulated as

\begin{equation}\label{expectation model}
\begin{aligned}
& \min _{\mathbf{X}} \  \mathbb{E}\left(\mathbf{h}_{j, b}^H \mathbf{X} \mathbf{X}^H \mathbf{h}_{j, b}\right)+\mu \mathbb{E}\left(\mathcal{D}\left(\mathbb{P}_1 \| \mathbb{P}_0\right)\right) \\
& \text { s.t. } \  \mathbf{X}^H \mathbf{X}=\frac{n}{p} \mathbf{I}_p
\end{aligned}
\end{equation}
where $\mu$ is a preset positive weight.

Since the probability density function (PDF) of $\mathbf{h}_{j,b}$ and the form of the random variable function $\mathcal{D}\left(\mathbb{P}_1 \| \mathbb{P}_0\right)$ is very complex, an exact analysis expression of $ \mathbb{E}\left(\mathcal{D}\left(\mathbb{P}_1 \| \mathbb{P}_0\right)\right)$ is mathematically intractable.

According to \cite{online-learning}, when the sample size is large, the arithmetic mean can be a good estimate of statistical mean. Let $g_1(\mathbf{X},\mathbf{h}_{j,b}) = \mathbf{h}_{j, b}^H \mathbf{X} \mathbf{X}^H \mathbf{h}_{j, b} $, $g_2(\mathbf{X},h_{a,w},\mathbf{h}_{j,w}) = \mathcal{D}\left(\mathbb{P}_1 \| \mathbb{P}_0\right) $, we have
\begin{subequations}
\begin{align}
& \mathbb{E}\left( \mathbf{h}_{j, b}^H \mathbf{X} \mathbf{X}^H \mathbf{h}_{j, b}\right)=\frac{1}{T} \label{E1 to g1} \sum_{t=1}^T g_1\left(\mathbf{X},\mathbf{h}_{j,b}\left[t\right]\right) \\
& \mathbb{E}\left(\mathcal{D}\left(\mathbb{P}_1 \| \mathbb{P}_0\right)\right)=\frac{1}{T} \sum_{t=1}^T g_2\left(\mathbf{X}, h_{a, w}[t], \mathbf{h}_{j, w}[t]\right)  \label{E2 to g2}
\end{align}
\end{subequations}
where $\mathbf{h}_{j,b}\left[t\right]$, $h_{a,w}\left[t\right]$ and $\mathbf{h}_{j,w}\left[t\right]$ respectively represent the $t$-th sample realization of the channel coefficients from Jammer to Bob, Alice to Willie, and Jammer to Willie, and $T$ represents the number of samples.

Specifically, $g_1(\mathbf{X},\mathbf{h}_{j,b}\left[t\right])$, $g_2\left(\mathbf{X},h_{a,w}\left[t\right],h_{j,w}\left[t\right]\right)$ is represented as follows
\begin{subequations}
\begin{align}
&g_1(\mathbf{X},\mathbf{h}_{j,b}\left[t\right]) = \mathbf{h}_{j, b}^H[t] \mathbf{X} \mathbf{X}^H \mathbf{h}_{j, b}[t]   \label{g1 expression}  \\ \notag
&g_2\left(\mathbf{X}, h_{a, w}[t], \mathbf{h}_{j, w}[t]\right)\\  \notag
&= \ln \left(P_a\left|h_{a, w}[t]\right|^2+P_j \mathbf{h}_{j, w}^H[t] \mathbf{X} \mathbf{X}^H \mathbf{h}_{j, w}[t]+\sigma_w^2\right)  \label{g2 expression} \\
&\ \ \  -\ln \left(P_j \mathbf{h}_{j, w}^H[t] \mathbf{X} \mathbf{X}^H \mathbf{h}_{j, w}[t]+\sigma_w^2\right)\\
&\ \ \ -\frac{P_a\left|h_{a, w}[t]\right|^2}{P_a\left|h_{a, w}[t]\right|^2+P_j \mathbf{h}_{j, w}^H[t] \mathbf{X} \mathbf{X}^H \mathbf{h}_{j, w}[t]+\sigma_w^2} \notag
\end{align}
\end{subequations}

According to \eqref{E1 to g1}, \eqref{E2 to g2}, \eqref{g1 expression} and \eqref{g2 expression}, the optimization model is transformed to
\begin{equation}\label{original problem}
\begin{aligned}
\hspace{-0.5cm}& \min _{\mathbf{X}}   \frac{1}{T} \sum_{t=1}^T\left(g_1\left(\mathbf{X},\mathbf{h}_{j,b}\left[t\right]\right)\!+\!\mu g_2\left(\mathbf{X}, h_{a, w}[t], \mathbf{h}_{j, w}[t]\right)\right) \\
& \text { s.t. }   \mathbf{X}^H \mathbf{X}=\frac{n}{p} \mathbf{I}_p
\end{aligned}
\end{equation}

By analyzing optimization problem \eqref{original problem}, we observe that the constraints of this optimization problem are non-convex, making it difficult to handle directly. If we employ traditional matrix reshaping operations to convert these constraints into equality constraints, it would destroy the structure of the original matrix variable $\mathbf{X}$ and significantly increase the complexity of the problem's solution. Additionally, the computational cost of the full gradient of the objective function scales linearly with the number of samples $T$, further adding to the complexity of the problem. Therefore, the key to solving the problem lies in how to leverage the structure of the unitary matrix constraint and simplify the computational cost of gradient computation at each iteration.

In summary, the difficulty in solving optimization problem \eqref{original problem} lies in handling the unitary matrix constraint and reducing the computational cost of gradient computation at each iteration, which will be discussed in the next section.

\section{PROPOSED ALGORITHM}

As mentioned in the previous section, solving the optimization problem \eqref{original problem} directly is extremely challenging due to the non-convex constraints. However, we aim to utilize the structure of the constraints to design optimization algorithms. Riemannian manifold optimization based on Riemannian geometry theory can effectively leverage the intrinsic geometric structure of the constraint, embedding it into the search space, and transforming the original problem into an unconstrained optimization problem on the manifold. Through analyzing the form of the constraint, it is found that the corresponding constraint is the complex Stiefel manifold. To address the issue of excessive computational cost for computing the full gradient of the objective function, we consider accelerating stochastic gradient descent methods. Specifically, in this paper, we use the R-SVRG (Riemannian Stochastic Variance Reduced Gradient) algorithm to solve the optimization problem. In this section, we first analyze the geometric structure of the optimization problem, then customize the R-SVRG algorithm on the complex Stiefel manifold to solve problem \eqref{original problem}.

\subsection{Analysis of The Structure of the Constraint}

The constraint $\mathbf{X}^H\mathbf{X}=\frac{n}{p}\mathbf{I}_p$ can be simplified to $\mathbf{X}^H\mathbf{X} = \mathbf{I}_p$, and then letting $\sqrt{\frac{n}{p}}\mathbf{X}^*$ be the solution to the original optimization problem \eqref{original problem}. Regarding the treatment of constraint $\mathbf{X}^H\mathbf{X} = \mathbf{I}_p$, we embed it into the search space, where the search space transitions from the original space $\mathbb{C}^{n \times p}$ to the set $\left\{\mathbf{X} \in \mathbb{C}^{n \times p}: \mathbf{X}^H \mathbf{X}=\mathbf{I}_p\right\}$. The manifold corresponding to this set is the complex Stiefel manifold \cite{complex-stiefel}, mathematically represent as $\operatorname{St}\left(p,n,\mathbb{C}\right)$. The original non-convex constrained optimization problem is transformed into an unconstrained problem
\begin{equation} \label{manifold problem}
\min _{\mathbf{X} \in \operatorname{St}\left(p,n,\mathbb{C}\right)} \  \frac{1}{T} \sum_{t=1}^T\left(g_1\left(\mathbf{X},\mathbf{h}_{j,b}\left[t\right]\right)+\mu g_2\left(\mathbf{X}, h_{a, w}[t], \mathbf{h}_{j, w}[t]\right)\right)
\end{equation}

\subsubsection{The Geometric Structure of the Complex Stiefel Manifold}

Complex matrix $\mathbf{X} = \operatorname{Re}\left(\mathbf{X}\right) + i\operatorname{Im}\left(\mathbf{X}\right) \in \mathbb{C}^{n \times p}$, $\operatorname{Re}\left(\mathbf{X}\right), \operatorname{Im}\left(\mathbf{X}\right) \in \mathbb{R}^{n \times p}$, can be expressed as a $2 n \times 2 p$
real matrix
\begin{equation}\label{complex matrix real expression}
\bar{\mathbf{X}}:=\left(\begin{array}{cc}\operatorname{Re}\left(\mathbf{X}\right) & \operatorname{Im}\left(\mathbf{X}\right) \\ -\operatorname{Im}\left(\mathbf{X}\right) & \operatorname{Re}\left(\mathbf{X}\right)\end{array}\right)
\end{equation}

The real representation $\operatorname{Stp}\left(p,n\right)$ of the complex Stiefel manifold $\operatorname{St}\left(p,n,\mathbb{C}\right)$ is \cite{complex-stiefel}
\begin{equation}\label{stp manifold}
\operatorname{Stp}\left(p,n\right) := \operatorname{St}\left(2p,2n,\mathbb{R}\right) \cap \mathcal{S P}\left(p,n\right)
\end{equation}
where $\bar{\mathbf{X}} \in \operatorname{Stp}\left(p,n\right)$, $\operatorname{St}\left(2p,2n,\mathbb{R}\right)$ is the real Stiefel manifold defined by $\operatorname{St}\left(2p,2n,\mathbb{R}\right) = \left\{\mathbf{Y} \in \mathbb{R}^{2n \times 2p}: \mathbf{Y}^T \mathbf{Y}=\mathbf{I}_{2p}\right\}$ and $\mathcal{S P}\left(p,n\right)$ is the quasi-symplectic set defined by $\mathcal{S P}(p, n)=\left\{\mathbf{Z} \in \mathbb{R}^{2 n \times 2 p} \mid \mathbf{Z} \mathbf{J}_p=\mathbf{J}_n \mathbf{Z}\right\}$, and $\mathbf{J}_p:=\left(\begin{array}{cc}0 & \mathbf{I}_p \\ -\mathbf{I}_p & 0\end{array}\right)$, $\mathbf{J}_n:=\left(\begin{array}{cc}0 & \mathbf{I}_n \\ -\mathbf{I}_n & 0\end{array}\right)$.

\begin{fact}\label{fact-complex-real}
For a complex matrix $\mathbf{X} \in \operatorname{St}\left(p,n,\mathbb{C}\right)$ is equivalent to $\bar{\mathbf{X}} \in \operatorname{St}\left(2p,2n,\mathbb{R}\right)$.
\end{fact}

{\it Proof:} Please see Appendix~\ref{fact1 proof}.

In the same way that the derivative of real-valued function provides a local linear approximation of the function, the tangent space provides a local vector space approximation of the manifold. The tangent space to $\operatorname{Stp}\left(p,n\right)$ at $ \bar{\mathbf{X}}$ is given by
\begin{equation}\label{Stp tangent space}
 T_{\bar{\mathbf{X}}}\operatorname{Stp}(p, n)
= \{(\bar{\xi}_{\bar{\mathbf{X}}} \in \mathcal{S} \mathcal{P}(p, n) \mid \left.\bar{\xi}_{\bar{\mathbf{X}}}^T \bar{\mathbf{X}}+\bar{\mathbf{X}}^T \bar{\xi}_{\bar{\mathbf{X}}}=0\right\}
\end{equation}

Because the embedding space of manifold $\operatorname{Stp}\left(p,n\right)$ is $\mathbb{R}^{2n \times 2p}$, $T_{\bar{\mathbf{X}}}(\operatorname{Stp}(p, n))$ is a subspace of $\mathbb{R}^{2n \times 2p}$, the Riemannian metric on manifold $\operatorname{Stp}\left(p,n\right)$ can be expressed by the inner product on the Euclidean space $\mathbb{R}^{2n \times 2p}$ as
\begin{equation}\label{Riemannian metric}
g_{\bar{\mathbf{x}}}(\bar{\xi}_{\bar{\mathbf{X}}}, \bar{\eta}_{\bar{\mathbf{X}}})=\frac{1}{2}\operatorname{tr}\left(\bar{\xi}_{\bar{\mathbf{X}}}^T \bar{\eta}_{\bar{\mathbf{X}}}\right)
\end{equation}
where $\bar{\xi}_{\bar{\mathbf{X}}}, \bar{\eta}_{\bar{\mathbf{X}}} \in T_{\bar{\mathbf{X}}}\operatorname{Stp}(p, n)$.

Then, the orthogonal complement space of the tangent space  $T_{\bar{\mathbf{X}}}(\operatorname{Stp}(p, n))$ is given by
\begin{equation}\label{Stp orthogonal complement space}
\left(T_{\bar{\mathbf{X}}} \operatorname{Stp}(p, n)\right)^{\perp}=\left\{\bar{\mathbf{X}} \mathbf{S}: \mathbf{S} \in S_{s y m}(2p)\right\}
\end{equation}
where $S_{s y m}(2p)$ is the set of $2 p \times 2 p$ dimensional real symmetric matrices.

{\it Proof:} The proof regarding the tangent space, Riemannian metric, and the orthogonal complement of the tangent space is provided in Appendix~\ref{complex stiefel proof}.

\subsubsection{Riemannian Gradient}
Given a smooth real-valued function $\bar{f}\left(\bar{\mathbf{X}}\right)$ on manifold $\operatorname{Stp}\left(p,n\right)$. The Riemannian gradient $\operatorname{grad}\bar{f}\left(\bar{\mathbf{X}}\right)$ of $\bar{f}\left(\bar{\mathbf{X}}\right)$ at $\bar{\mathbf{X}}$ represents the steepest ascent direction at point $\bar{\mathbf{X}}$ denoted as
\begin{equation}\label{gradient expression}
\operatorname{grad} \bar{f}(\bar{\mathbf{X}})=\mathrm{P}_{\bar{\mathbf{X}}} \nabla \bar{f}(\bar{\mathbf{X}})
\end{equation}
where $\nabla \bar{f}(\bar{\mathbf{X}})$ is the Euclidean gradient of $\bar{f}\left(\bar{\mathbf{X}}\right)$, and $\mathrm{P}_{\bar{\mathbf{X}}} \nabla \bar{f}(\mathbf{X})$ denotes the orthogonal projective from $\nabla \bar{f}(\bar{\mathbf{X}})$ to $T_{\bar{\mathbf{X}}}\operatorname{Stp}\left(p,n\right)$.

\begin{lemma}\label{lemma-Stp-gradient}
For manifold $\operatorname{Stp}\left(p,n\right)$, the Riemannian gradient is given by
\begin{equation}\label{Stp rgrad}
\operatorname{grad} \bar{f}(\bar{\mathbf{X}})\!=\!\left(\mathbf{I}\!-\!\bar{\mathbf{X}} \bar{\mathbf{X}}^T\right) \nabla \bar{f}(\bar{\mathbf{X}})\!+\!\bar{\mathbf{X}}  \text{skew}\left(\bar{\mathbf{X}}^T \nabla \bar{f}(\bar{\mathbf{X}})\right)
\end{equation}
where  $$\text {skew}\left(\bar{\mathbf{X}}^T \nabla \bar{f}(\bar{\mathbf{X}})\right)=\frac{1}{2}\left( \bar{\mathbf{X}}^T \nabla \bar{f}\left(\bar{\mathbf{X}}\right)- \left(\bar{\mathbf{X}}^T \nabla \bar{f}\left(\bar{\mathbf{X}}\right)\right)^T\right)$$

\end{lemma}

{\it Proof:} Please see Appendix~\ref{lemma1 proof}.

\begin{lemma}\label{lemma-complex-stiefel-gradient}
For the real-valued complex variable function $f\left(\mathbf{X}\right)$ corresponding to $\bar{f}\left(\bar{\mathbf{X}}\right)$, the Riemannian gradient of manifold $\operatorname{S t}\left(p,n,\mathbb{C}\right)$ can be derived form {\it Lemma~\ref{lemma-Stp-gradient}} as
\begin{equation}\label{complex stiefel rgrad}
\operatorname{grad} f(\mathbf{X})=\left(\mathbf{I}-\mathbf{X} \mathbf{X}^H\right) \nabla f(\mathbf{X})+\mathbf{X}  \text{skew}\left(\mathbf{X}^H \nabla f(\mathbf{X})\right)
\end{equation}
where $\nabla f(\mathbf{X})$ is the Euclidean gradient of $f\left(\mathbf{X}\right)$ and $$\text {skew}\left(\mathbf{X}^H \nabla f(\mathbf{X})\right)=\frac{1}{2}\left( \mathbf{X}^H \nabla f\left(\mathbf{X}\right)- \left(\mathbf{X}^H \nabla f\left(\mathbf{X}\right)\right)^H\right)$$
\end{lemma}

{\it Proof:} Please see Appendix~\ref{lemma2 proof}.

\subsubsection{Retraction}
The extension of line search methods from Euclidean space to manifolds operates as follow: Select a tangent vector $\bar{\xi}$ at point $\bar{\mathbf{X}}$ on the manifold $\operatorname{Stp}\left(p,n\right)$, and search along a curve $\gamma \left(t\right)$ on the manifold, where the tangent vector at $\gamma \left(0\right)$ is  $\bar{\xi}$. Since performing this operation directly on curve $\gamma \left(0\right)$ is challenging, the concept of retraction is introduced to enable searching along the tangent vector direction on the manifold.

\begin{figure}[!t]
\centering
\includegraphics[width=3.6in]{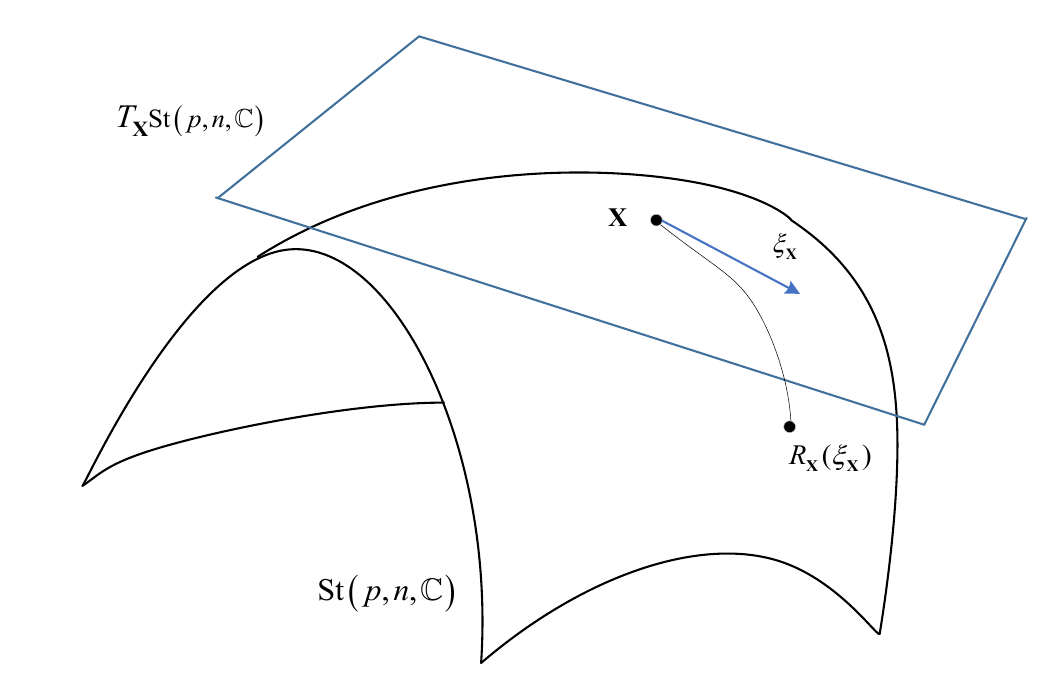}
\caption{Illustration of the retraction on the complex Stiefel manifold.}
\label{retraction illustrate}
\end{figure}

A retraction at $\bar{\mathbf{X}}$, denoted by $R_{\bar{\mathbf{X}}}$ is a mapping from $T_{\bar{\mathbf{X}}}\operatorname{Stp}\left(p,n\right)$ to $\operatorname{Stp}\left(p,n\right)$ with a local rigidity condition that preserves gradients at $\bar{\mathbf{X}}$. See Fig.~\ref{retraction illustrate}.
\begin{lemma}\label{lemma-Stp-retraction}
The retraction on $\operatorname{Stp}\left(p,n\right)$ is given by
\begin{equation}\label{Stp retraction}
R_{\bar{\mathbf{X}}}(\bar{\xi}_{\bar{\mathbf{X}}})=\phi\left(R_{\phi^{-1}(\bar{\mathbf{X}})}\left(\phi^{-1}(\bar{\xi}_{\bar{\mathbf{X}}})\right)\right)=\phi\left(R_{\mathbf{X}}(\xi_{\mathbf{X}})\right)
\end{equation}
where $\bar{\mathbf{X}} \in \operatorname{Stp}\left(p,n\right)$, $\bar{\xi}_{\bar{\mathbf{X}}} \in T_{\bar{\mathbf{X}}}\operatorname{Stp}\left(p,n\right)$, $\mathbf{X} \in \operatorname{St}\left(p,n,\mathbb{C}\right)$, $\xi_{\mathbf{X}} \in T_{\mathbf{X}}\operatorname{St}\left(p,n,\mathbb{C}\right)$, $\phi \left(\mathbf{X}\right)=\bar{\mathbf{X}}$, $R_{\mathbf{X}}\left(\xi_{\mathbf{X}}\right)=\operatorname{qf}\left(\mathbf{X}+\xi_\mathbf{X}\right)$ and the $\operatorname{qf}\left(\cdot\right)$ denotes the Q-factor of the complex QR decomposition.
\end{lemma}

{\it Proof:} Please see Appendix~\ref{lemma3 proof}.

According to {\it Lemma \ref{lemma-Stp-retraction}}, the retraction $R_{\mathbf{X}}$ on $\operatorname{St}\left(p,n,\mathbb{C}\right)$ is given by
\begin{equation}\label{complex stiefel retraction}
R_{\mathbf{X}}\left(\xi_\mathbf{X}\right)=\operatorname{qf}\left(\mathbf{X}+\xi_\mathbf{X}\right)
\end{equation}

\subsubsection{Vector Transport}
In the algorithms mentioned in the next subsection, computing the update direction requires Riemannian gradients not only from the current point but also from other points. Since the tangent spaces at different points are distinct vector spaces, they cannot be directly added. Addition of Riemannian gradients at different points is achieved by introducing the concept of vector transport.

For simplify, the vector transport $\mathcal{T}$ on manifold $\operatorname{St}\left(p,n,\mathbb{C}\right)$ accomplishes the transfer of tangent vector $\xi_{\mathbf{X}}$ from the tangent space $T_{\mathbf{X}}\operatorname{St}\left(p,n,\mathbb{C}\right)$ at point $\mathbf{X}$ to the tangent space $T_{R_{\mathbf{X}}\left(\eta_{\mathbf{X}}\right)}\operatorname{St}\left(p,n,\mathbb{C}\right)$ at point $R_{\mathbf{X}}\left(\eta_{\mathbf{X}}\right)$, along the update of tangent vector $\eta_{\mathbf{X}}$. The specific operation is given by
\begin{equation}\label{vector transport define}
\mathcal{T}_{\eta_{\mathbf{X}}} \xi_{\mathbf{X}}=\mathrm{P}_{R_{\mathbf{X}}\left(\eta_{\mathbf{X}}\right)} \xi_{\mathbf{X}}
\end{equation}
where $\mathrm{P}_{R_{\mathbf{X}}}$ represents the orthogonal projection onto space $T_{R_{\mathbf{X}}\left(\eta_{\mathbf{X}}\right)}\operatorname{St}\left(p,n,\mathbb{C}\right)$.

It's easy to see here that the vector transport operator is a linear operator, i.e.
\begin{equation} \label{vector transport linear}
\mathcal{T}_{\eta_{\mathbf{x}}}\left(a \xi_{\mathbf{X}}+b \varsigma_{\mathbf{X}}\right)=a \mathcal{T}_{\eta_{\mathbf{x}}}\left(\xi_{\mathbf{X}}\right)+b \mathcal{T}_{\eta_{\mathbf{x}}}\left(\varsigma_{\mathbf{X}}\right)
\end{equation}
where $a,b$ are scalar, $\eta_{\mathbf{x}}, \varsigma_{\mathbf{X}}, \xi_{\mathbf{X}} \in T_{\mathbf{X}}\operatorname{St}\left(p,n,\mathbb{C}\right)$.

\begin{lemma}\label{lemma-vetor-transport}
According to {\it Lemma \ref{lemma-complex-stiefel-gradient}}, the vector transport on $\operatorname{St}\left(p,n,\mathbb{C}\right)$ is defined by
\begin{equation}\label{complex stiefel vector transport}
\mathcal{T}_{\eta_{\mathbf{X}}} \xi_{\mathbf{X}}=\left(\mathbf{I}-\mathbf{Y} \mathbf{Y}^H\right) \xi_{\mathbf{X}}+\mathbf{Y} \operatorname{skew}\left(\mathbf{Y}^H \xi_{\mathbf{X}}\right)
\end{equation}
where $\mathbf{Y}=R_{\mathbf{X}}\left(\eta_{\mathbf{X}}\right)$.
\end{lemma}

This subsection analyzes the geometric structure of the constraint in optimization problems to transform the originally non-convex constrained optimization problem \eqref{original problem} into an unconstrained problem on the complex Stiefel manifold for solution, and provides the expression of the required operators. The next subsection customizes optimization algorithms for solving optimization problem \eqref{manifold problem} based on the objective function.

\subsection{Proposed Algorithm}
In this subsection, we propose an algorithm to solve optimization problem \eqref{manifold problem}, which has the following three advantages:
\begin{itemize}
\item Lower computational cost.
\item Customized algorithms structure.
\item Theoretical convergence guarantee.
\end{itemize}

For optimization problem \eqref{manifold problem}, the iterative algorithm based on the full gradient requires computation of gradients for $T$ iterations. Given the potentially large value of $T$, this leads to substantial computational overhead in each iteration. In this subsection, we adopt the stochastic gradient descent approach to reduce computational cost. Specifically, we employ the Riemannian stochastic variance reduced gradient (R-SVRG) algorithm \cite{R-SVRG}. Furthermore, we iteratively solve optimization problem \eqref{manifold problem} by customizing the R-SVRG algorithm on the complex Stiefel manifold $\operatorname{St}\left(p,n,\mathbb{C}\right)$.

Here, let's first review the SVRG algorithm in Euclidean space \cite{SVRG}. Addressing the issue of large computational overhead for gradient computation in each iteration as mentioned earlier, an alternative method is the stochastic gradient descent (SGD) algorithm. In each iteration, the negative gradient direction $- \nabla f_{t_q}\left(\mathbf{X}\right)$ of a randomly selected function $f_{t_q}\left(\mathbf{X}\right)$ is chosen as the descent direction. The update criterion at time $q = 1,2,...$ is expressed as follows
\begin{equation}\label{SGD update}
\mathbf{X}^{q} = \mathbf{X}^{q-1}-\alpha_q \nabla f_{t_q}\left(\mathbf{X}^{q-1}\right)
\end{equation}
where $t_q$ is obtained by randomly selecting from $\{1,...,T\}$ at the $q$-th iteration.

Although the SGD algorithm exhibits randomness, on average, the result of each iteration is equivalent to that of full gradient descent, and each iteration significantly reduces computational overhead compared to full gradient descent. However, the random selection of gradients during the iteration process introduces variance. When the variance is high, the algorithm may exhibit large fluctuations in gradients during execution, leading to instability and slowing convergence.

The SVRG algorithm utilizes variance reduction techniques to accelerate the convergence of SGD. It employs a two-layer loop, where the superscript $``l"$ denotes the current iteration of the outer loop, the subscript $``q"$ represents the current iteration of the inner loop within the $l$-th outer loop, and there are $m_l$ iteration in each outer loop. The random gradient subscript $t_q^l$ is obtained by randomly selecting from $\{1,...,T\}$. The gradient for each iteration is given by
\begin{equation}\label{SVRG gradient}
\varsigma_q^l=\nabla f_{t_q^l}\left(\mathbf{X}_{q-1}^l\right)-\nabla f_{t_q^l}\left(\widetilde{\mathbf{X}}^{l-1}\right)+\nabla f\left(\widetilde{\mathbf{X}}^{l-1}\right)
\end{equation}
where $\widetilde{\mathbf{X}}^{l-1}=\mathbf{X}^{l-1}_{m_{l-1}}$, and the initial point for the $l$-th outer loop is $\mathbf{X}_0^{l}=\widetilde{\mathbf{X}}^{l-1}$. The update criterion for the SVRG algorithm is as follows
\begin{equation}\label{SVRG update}
\mathbf{X}_q^l=\mathbf{X}_{q-1}^l-\alpha_q^l\varsigma_q^l
\end{equation}
where $\alpha_q^l$ is the step size in the $q$-th iteration of the $l$-th outer loop.

The R-SVRG algorithm is an extension of the aforementioned Euclidean SVRG algorithm into Riemannian manifold. Specifically, R-SVRG retains point $\widetilde{\mathbf{X}}^{l-1}\in \operatorname{St}\left(p,n,\mathbb{C}\right)$ after $m_{l-1}$ iterations in the $(l-1)$-th outer loop, computes the full gradient $\operatorname{grad}f\left(\widetilde{\mathbf{X}}^{l-1}\right)$, and calculates the Riemannian random gradient $\operatorname{grad}f_{t_q^l}\left(\widetilde{\mathbf{X}}^{l-1}\right)$ to reduce the variance of the stochastic gradient $\operatorname{grad}f_{t_q^l}\left(\mathbf{X}_{q-1}^l\right)$.

\begin{figure}[!t]
\centering
\includegraphics[width=3.6in]{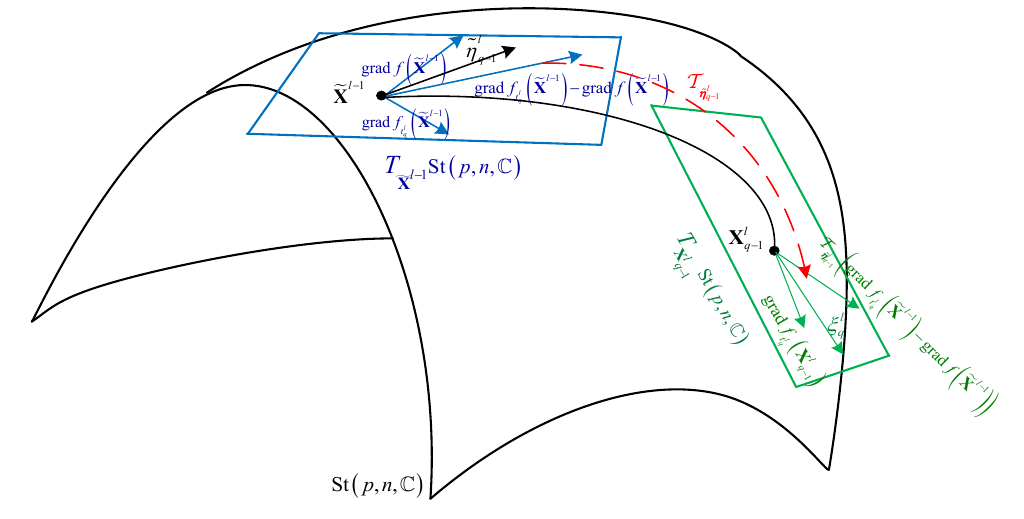}
\caption{The gradient of R-SVRG algorithm on the complex Stiefel manifold.}
\label{R-SVRG Gradient}
\end{figure}

Drawing from the gradient computation approach in Euclidean SVRG algorithm, in R-SVRG. the gradients is a linear combination of $\operatorname{grad}f_{t_q^l}\left(\mathbf{X}_{q-1}^l\right)$, $-\operatorname{grad}f_{t_q^l}\left(\widetilde{\mathbf{X}}^{l-1}\right)$ and $\operatorname{grad}f\left(\widetilde{\mathbf{X}}^{l-1}\right)$. However, due to $\operatorname{grad}f_{t_q^l}\left(\mathbf{X}_{q-1}^l\right) \in T_{\mathbf{X}_{q-1}^l}\operatorname{St}\left(p,n,\mathbb{C}\right)$; $-\operatorname{grad}f_{t_q^l}\left(\widetilde{\mathbf{X}}^{l-1}\right), \operatorname{grad}f\left(\widetilde{\mathbf{X}}^{l-1}\right) \in T_{\widetilde{\mathbf{X}}^{l-1}}\operatorname{St}\left(p,n,\mathbb{C}\right) $ and $T_{\mathbf{X}_{q-1}^l}\operatorname{St}\left(p,n,\mathbb{C}\right)$, $T_{\widetilde{\mathbf{X}}^{l-1}}\operatorname{St}\left(p,n,\mathbb{C}\right)$ being two separate vector spaces, direct addition operation is not feasible. Here, we utilize the vector transport operator mentioned in the previous subsection to achieve addition of tangent vectors between different tangent spaces. Specifically, we first transport $\operatorname{grad}f\left(\widetilde{\mathbf{X}}^{l-1}\right)-\operatorname{grad}f_{t_q^l}\left(\widetilde{\mathbf{X}}^{l-1}\right)$ from space $T_{\widetilde{\mathbf{X}}^{l-1}}\operatorname{St}\left(p,n,\mathbb{C}\right)$ to the tangent space $T_{\mathbf{X}_{q-1}^l}\operatorname{St}\left(p,n,\mathbb{C}\right)$  at point $\mathbf{X}_{q-1}^l$, and then add it to $\operatorname{grad}f_{t_q^l}\left(\mathbf{X}_{q-1}^l\right)$. The specific operation is illustrated in Fig.~\ref{R-SVRG Gradient}.

The gradient of each iteration of R-SVRG is
\begin{equation}\label{R-SVRG gradient}
\begin{aligned}
&\xi_q^l= \\
&\operatorname{grad}\! f_{t_q^l}\left(\mathbf{X}_{q-1}^l\right)\!-\!\mathcal{T}_{\tilde{\eta}_{q-1}^l}\left(\!\operatorname{grad} f_{t_q^l}\left(\widetilde{\mathbf{X}}^{l-1}\right)\!-\!\operatorname{grad} f\left(\widetilde{\mathbf{X}}^{l-1}\right)\!\right)
\end{aligned}
\end{equation}

The update rule for R-SVRG is as follows
\begin{equation}\label{R-SVRG update}
\mathbf{X}_q^l=R_{\mathbf{X}_{q-1}^l}\left(-\alpha_{q-1}^l \xi_q^l\right)
\end{equation}

To specifically solve optimization problem \eqref{manifold problem}, we first compute the Euclidean gradients of the objective function $f\left(\mathbf{X}\right) =  \frac{1}{T} \sum_{t=1}^T\left(g_1\left(\mathbf{X},\mathbf{h}_{j,b}\left[t\right]\right)+\mu g_2\left(\mathbf{X}, h_{a, w}[t], \mathbf{h}_{j, w}[t]\right)\right) $, denoted as $\nabla f_{t_q^l}\left(\mathbf{X}_{q-1}^l\right), \nabla f_{t_{q}^{l}}\left(\widetilde{\mathbf{X}}^{l-1}\right)$, $\nabla f\left(\widetilde{\mathbf{X}}^{l-1}\right)$. The specific expressions are as \eqref{Euclidean sample gradient}, \eqref{Euclidean sample gradient save}, \eqref{Euclidean full gradient save}.

\begin{figure*}
\begin{equation} \label{Euclidean sample gradient}
\begin{aligned}
\nabla f_{t_q^l}\left(\mathbf{X}_{q-1}^l\right) =     & 2\left(\mathbf{h}_{j,b}\left[t_q^l\right]\mathbf{h}_{j,b}^H\left[t_q^l\right]\mathbf{X}_{q-1}^l+ \frac{P_j\mathbf{h}_{j,w}\left[t_q^l\right]\mathbf{h}_{j,w}^H\left[t_q^l\right]\mathbf{X}_{q-1}^l}{P_a\left|h_{a, w}\left[t_q^l\right]\right|^2+P_j \mathbf{h}_{j, w}^H\left[t_q^l\right] \mathbf{X}_{q-1}^l \left(\mathbf{X}_{q-1}^l\right)^H \mathbf{h}_{j, w}\left[t_q^l\right]+\sigma_w^2}\right. \\
&\left.-\frac{P_j\mathbf{h}_{j,w}\left[t_q^l\right]\mathbf{h}_{j,w}^H\left[t_q^l\right]\mathbf{X}_{q-1}^l}{P_j \mathbf{h}_{j, w}^H\left[t_q^l\right] \mathbf{X}_{q-1}^l \left(\mathbf{X}_{q-1}^l\right)^H \mathbf{h}_{j, w}\left[t_q^l\right]+\sigma_w^2}+\frac{P_a\left|h_{a,w}\left[t_q^l\right]\right|^2P_j\mathbf{h}_{j,w}
\left[t_q^l\right]\mathbf{h}_{j,w}^H\left[t_q^l\right]\mathbf{X}_{q-1}^l}{\left|P_a\left|h_{a,w}\left[t_q^l\right]\right|^2+P_j \mathbf{h}_{j, w}^H\left[t_q^l\right] \mathbf{X}_{q-1}^l \left(\mathbf{X}_{q-1}^l\right)^H \mathbf{h}_{j, w}\left[t_q^l\right]+\sigma_w^2\right|^2}\right)
\end{aligned}
\end{equation}
\begin{equation}
\begin{aligned} \label{Euclidean sample gradient save}
\nabla f_{t_q^l}\left(\widetilde{\mathbf{X}}^{l-1}\right) = & 2\left(\mathbf{h}_{j,b}\left[t_q^l\right]\mathbf{h}_{j,b}^H\left[t_q^l\right]\widetilde{\mathbf{X}}^{l-1}+ \frac{P_j\mathbf{h}_{j,w}\left[t_q^l\right]\mathbf{h}_{j,w}^H\left[t_q^l\right]\widetilde{\mathbf{X}}^{l-1}}{P_a\left|h_{a, w}\left[t_q^l\right]\right|^2+P_j \mathbf{h}_{j, w}^H\left[t_q^l\right] \widetilde{\mathbf{X}}^{l-1} \left(\widetilde{\mathbf{X}}^{l-1}\right)^H \mathbf{h}_{j, w}\left[t_q^l\right]+\sigma_w^2}\right.\\
&\left.-\frac{P_j\mathbf{h}_{j,w}\left[t_q^l\right]\mathbf{h}_{j,w}^H\left[t_q^l\right]\widetilde{\mathbf{X}}^{l-1}}{P_j \mathbf{h}_{j, w}^H\left[t_q^l\right] \widetilde{\mathbf{X}}^{l-1} \left(\widetilde{\mathbf{X}}^{l-1}\right)^H \mathbf{h}_{j, w}\left[t_q^l\right]+\sigma_w^2}+\frac{P_a\left|h_{a,w}\left[t_q^l\right]\right|^2P_j\mathbf{h}_{j,w}
\left[t_q^l\right]\mathbf{h}_{j,w}^H\left[t_q^l\right]\widetilde{\mathbf{X}}^{l-1}}{\left|P_a\left|h_{a,w}\left[t_q^l\right]\right|^2+P_j \mathbf{h}_{j, w}^H\left[t_q^l\right] \widetilde{\mathbf{X}}^{l-1} \left(\widetilde{\mathbf{X}}^{l-1}\right)^H \mathbf{h}_{j, w}\left[t_q^l\right]+\sigma_w^2\right|^2}\right)
\end{aligned}
\end{equation}
\begin{equation}\label{Euclidean full gradient save}
\begin{aligned}
\nabla f\left(\widetilde{\mathbf{X}}^{l-1}\right) =  &\frac{2}{T }\sum_{t=1}^T\left(\mathbf{h}_{j,b}\left[t\right]\mathbf{h}_{j,b}^H\left[t\right]\widetilde{\mathbf{X}}^{l-1}+ \frac{P_j\mathbf{h}_{j,w}\left[t\right]\mathbf{h}_{j,w}^H\left[t\right]\widetilde{\mathbf{X}}^{l-1}}{P_a\left|h_{a, w}\left[t\right]\right|^2+P_j \mathbf{h}_{j, w}^H\left[t\right] \widetilde{\mathbf{X}}^{l-1} \left(\widetilde{\mathbf{X}}^{l-1}\right)^H \mathbf{h}_{j, w}\left[t\right]+\sigma_w^2}\right.\\
&\left.-\frac{P_j\mathbf{h}_{j,w}\left[t\right]\mathbf{h}_{j,w}^H\left[t\right]\widetilde{\mathbf{X}}^{l-1}}{P_j \mathbf{h}_{j, w}^H\left[t\right] \widetilde{\mathbf{X}}^{l-1} \left(\widetilde{\mathbf{X}}^{l-1}\right)^H \mathbf{h}_{j, w}\left[t\right]+\sigma_w^2}+\frac{P_a\left|h_{a,w}\left[t\right]\right|^2P_j\mathbf{h}_{j,w}
\left[t\right]\mathbf{h}_{j,w}^H\left[t\right]\widetilde{\mathbf{X}}^{l-1}}{\left|P_a\left|h_{a, w}\left[t\right]\right|^2+P_j \mathbf{h}_{j, w}^H\left[t\right] \widetilde{\mathbf{X}}^{l-1} \left(\widetilde{\mathbf{X}}^{l-1}\right)^H \mathbf{h}_{j, w}\left[t\right]+\sigma_w^2\right|^2}\right)
\end{aligned}
\end{equation}
\hrulefill
\vspace*{4pt}
\end{figure*}

According to \eqref{complex stiefel rgrad}, the Riemannian gradients $\operatorname{grad}f_{t_q^l}\left(\mathbf{X}_{q-1}^l\right)$, $\operatorname{grad}f_{t_q^l}\left(\widetilde{\mathbf{X}}^{l-1}\right)$ and $\operatorname{grad}f\left(\widetilde{\mathbf{X}}^{l-1}\right)$ are respectively
\begin{equation}\label{Riemannian sample gradient}
\begin{aligned}
\operatorname{grad}f_{t_q^l}\left(\mathbf{X}_{q-1}^l\right)=&\left(\mathbf{I}-\mathbf{X}_{q-1}^l \left(\mathbf{X}_{q-1}^l\right)^H\right) \nabla f_{t_q^l}\left(\mathbf{X}_{q-1}^l\right)+ \\
&\mathbf{X}_{q-1}^l  \text{skew}\left(\left(\mathbf{X}_{q-1}^l\right)^H \nabla f_{t_q^l}\left(\mathbf{X}_{q-1}^l\right)\right)
\end{aligned}
\end{equation}
\begin{equation}\label{Riemannian sample gradient save}
\begin{aligned}
\operatorname{grad}f_{t_q^l}\left(\widetilde{\mathbf{X}}^{l-1}\right)=&\left(\mathbf{I}-\widetilde{\mathbf{X}}^{l-1} \left(\widetilde{\mathbf{X}}^{l-1}\right)^H\right) \nabla f_{t_q^l}\left(\widetilde{\mathbf{X}}^{l-1}\right)+ \\
&\widetilde{\mathbf{X}}^{l-1}  \text{skew}\left(\left(\widetilde{\mathbf{X}}^{l-1}\right)^H \nabla f_{t_q^l}\left(\widetilde{\mathbf{X}}^{l-1}\right)\right)
\end{aligned}
\end{equation}
\begin{equation}\label{Riemannian full gradient save}
\begin{aligned}
\operatorname{grad}f\left(\widetilde{\mathbf{X}}^{l-1}\right)=&\left(\mathbf{I}-\widetilde{\mathbf{X}}^{l-1} \left(\widetilde{\mathbf{X}}^{l-1}\right)^H\right) \nabla f\left(\widetilde{\mathbf{X}}^{l-1}\right)+ \\
&\widetilde{\mathbf{X}}^{l-1}  \text{skew}\left(\left(\widetilde{\mathbf{X}}^{l-1}\right)^H \nabla f\left(\widetilde{\mathbf{X}}^{l-1}\right)\right)
\end{aligned}
\end{equation}

Moreover, according to the vector transport operator \eqref{complex stiefel vector transport}, we have
\begin{equation}\label{vector transport sample}
\begin{aligned}
&\mathcal{T}_{\widetilde{\eta}_q^l}\left(\operatorname{grad}f_{t_q^l}\left(\widetilde{\mathbf{X}}^{l-1}\right)\right) =   \\ &\left(\mathbf{I}-\mathbf{X}_{q-1}^l\left(\mathbf{X}_{q-1}^l\right)^H\right)\operatorname{grad} f_{t_q^l}\left(\widetilde{\mathbf{X}}^{l-1}\right)+ \\
&\mathbf{X}_{q-1}^l\operatorname{skew}\left(\left(\mathbf{X}_{q-1}^l\right)^H\operatorname{grad}f_{t_q^l}\left(\widetilde{\mathbf{X}}^{l-1}\right)\right)
\end{aligned}
\end{equation}

\begin{equation}\label{vector transport full}
\begin{aligned}
&\mathcal{T}_{\widetilde{\eta}_q^l}\left(\operatorname{grad}f\left(\widetilde{\mathbf{X}}^{l-1}\right)\right) =   \\ &\left(\mathbf{I}-\mathbf{X}_{q-1}^l\left(\mathbf{X}_{q-1}^l\right)^H\right)\operatorname{grad} f\left(\widetilde{\mathbf{X}}^{l-1}\right)+ \\
&\mathbf{X}_{q-1}^l\operatorname{skew}\left(\left(\mathbf{X}_{q-1}^l\right)^H\operatorname{grad}f\left(\widetilde{\mathbf{X}}^{l-1}\right)\right)
\end{aligned}
\end{equation}

Since the vector transport operator is a linear operator, from \eqref{vector transport linear}, the gradient \eqref{R-SVRG gradient} for each iteration of R-SVRG can be written as
\begin{equation}\label{Riemannian gradient complex stiefel}
\begin{aligned}
\xi_q^l= &\operatorname{grad} f_{t_q^l}\left(\mathbf{X}_{q-1}^l\right)-\mathcal{T}_{\tilde{\eta}_{q-1}^l}\left(\operatorname{grad} f_{t_q^l}\left(\widetilde{\mathbf{X}}^{l-1}\right)\right)+\\
&\mathcal{T}_{\tilde{\eta}_{q-1}^l}\left(\operatorname{grad} f\left(\widetilde{\mathbf{X}}^{l-1}\right)\right)
\end{aligned}
\end{equation}

Then, the update criterion for variable $\mathbf{X}$ at each iteration is
\begin{equation}\label{complex stiefel update}
\mathbf{X}_q^l = \operatorname{qf}\left(\mathbf{X}_{q-1}^l-\alpha_{q-1}^l\xi_q^l\right)
\end{equation}

In summary, the R-SVRG algorithm process for optimization problem \eqref{manifold problem} is provided in Table \ref{proposed R-SVRG algorithm}.

\begin{table}[t]
\renewcommand \arraystretch{1.2}
\caption{The R-SVRG algorithm }
\label{proposed R-SVRG algorithm}
\centering
\begin{tabular}{l}
 \hline\hline
 {\bf Parameters:} Update frequency $m_l > 0$ and sequence $\{\alpha_q^l\}$ of \\ positive step sizes. \\
 {\bf Initialize:} $\widetilde{\mathbf{X}}^0$ \\
 {\bf Iterate:} for $l = 1,2,...$ \\
  \hspace{0.2cm} Calculate the full Riemannian gradient $\operatorname{grad}f\left(\widetilde{\mathbf{X}}^{l-1}\right)$ via \eqref{Euclidean full gradient save}, \eqref{Riemannian full gradient save}. \\
  \hspace{0.2cm} Store $\mathbf{X}_0^l = \widetilde{\mathbf{X}}^{l-1}$. \\
  \hspace{0.2cm} {\bf Iterate:} for $q = 1,2,...,m_l$ \\
  \hspace{0.4cm} Choose $t_q^l \in \{1,2,...,T\} $ uniformly at random.\\
  \hspace{0.4cm} Calculate the modified Riemannian stochastic gradient $\xi_q^l$ as \eqref{Riemannian gradient complex stiefel}\\
  \hspace{0.4cm} via \eqref{Euclidean sample gradient}-\eqref{vector transport full}.\\
  \hspace{0.4cm} Update $\mathbf{X}_q^l$ from $\mathbf{X}_{q-1}^l$ as \eqref{complex stiefel update}.\\
  \hspace{0.2cm} {\bf end}\\
  \hspace{0.2cm} set $\widetilde{\mathbf{X}}^l = \mathbf{X}_{m_l}^l$.\\
  {\bf Until} some preset termination criterion is satisfied.\\
  {\bf end} \\
 \hline\hline
\end{tabular}
\end{table}

\section{CONVERGENCE ANALYSIS}

We have the following theorem to show convergence properties of the proposed R-SVRG algorithm in Table \ref{proposed R-SVRG algorithm}.\\
\begin{theorem}\label{convergence theorem}
If the sequence $\{\alpha_q^l\}$ of positive step sizes satisfies
\begin{equation}\label{stepsize condition 1}
\sum_{l=1}^{\infty}\sum_{q=0}^{m_l}\left(\alpha_q^l\right)^2 < \infty
\end{equation}
\begin{equation}\label{stepsize condition 2}
\sum_{l=1}^{\infty}\sum_{q=0}^{m_l}\alpha_q^l = \infty
\end{equation}
Then, $\{f\left(\mathbf{X}_q^l\right)\}$ converges a.s. and $\operatorname{grad}f\left(\mathbf{X}_q^l\right) \rightarrow 0$ a.s. The proposed R-SVRG algorithm is convergent.
\end{theorem}

{\it Proof:} Please see Appendix G.

{\it Remark}: {\it Theorem \ref{convergence theorem}} indicates that the proposed R-SVRG algorithm on the complex Stiefel manifold theoretically converges to a static point of the optimization problem \eqref{manifold problem} under the conditions \eqref{stepsize condition 1} and \eqref{stepsize condition 2}. Here, it should be noted that condition \eqref{stepsize condition 1} ensures that the iteration step size decays to $0$, and condition \eqref{stepsize condition 2} ensures that the decay rate of the step size is not too fast. We can easily observe that the harmonic series satisfies the above two properties. Therefore, we choose $\alpha_q^l=\alpha_0\left(1+\alpha_0 \lambda\left\lfloor k_1 / m_l\right\rfloor\right)^{-1}$, where  $\alpha_0, \lambda>0$, $\alpha_0$ is the initial step size, $k_1$ is the total number of iterations depending on $l$ and $q$, and $\lfloor\cdot\rfloor$ denotes the floor function.

\section{NUMERICAL RESULTS}
In this section, we present numerical results to assess the performance of the proposed algorithm design with artificial noise in the scenario of covert communication aided by multiple friendly nodes. To demonstrate the effectiveness of our designed artificial noise signals, we compare this approach with a traditional Gaussian artificial noise strategy, where Gaussian signals are used as artificial noise. Without other statements, the parameters are set as follows: $P_a = 15$dBm, $P_j = 10$dBm, $\sigma_b^2 = \sigma_w^2 = 0$dBm, $T = 10000$, $\alpha_0 = 10^{-3}$, $m_l = 5T$, $\lambda = 20$. In what follows, we first simulate to demonstrate the convergence of our proposed algorithm. Then, we focus on investigating the impact of various key parameters under the condition of imperfect CSI considered in this paper on the covert transmission rate and covert performance.
\begin{figure}[!t]
\centering
\includegraphics[width=3.6in]{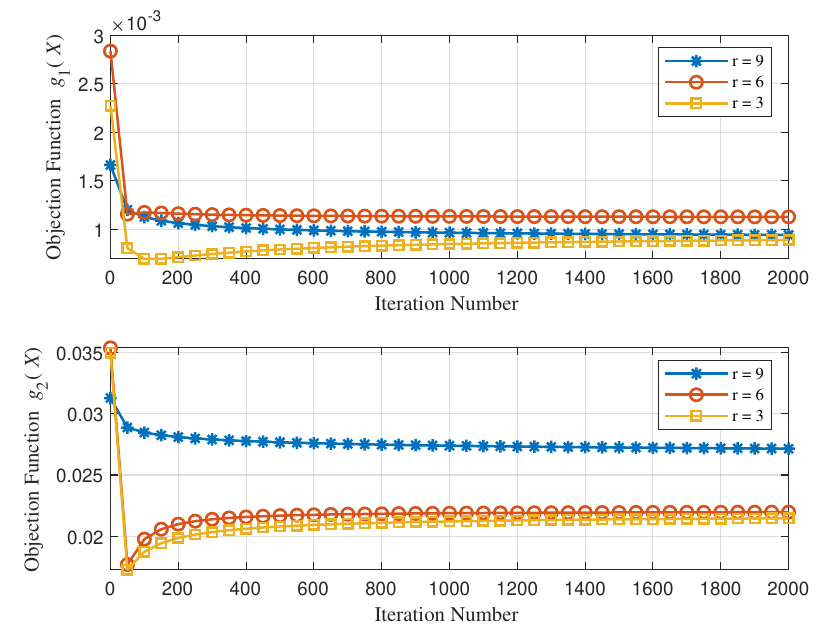}
\caption{$g_1\left(\mathbf{X}\right)$ and $g_2\left(\mathbf{X}\right)$ vs. iteration number.}
\label{obj convergence}
\end{figure}

Fig.~\ref{obj convergence} plots the performance curves of objective function $g_1\left(\mathbf{X}\right)$ and $g_2\left(\mathbf{X}\right)$ versus the iteration number for our proposed R-SVRG algorithms under three different scenarios: $n = 18, p = 9$; $n = 18, p = 6$ and $ n = 18, p = 3$. From the figure, it can be observed that the objective function $g_1\left(\mathbf{X}\right)$ converges in all three cases. Due to the stochastic nature of the R-SVRG algorithm, fluctuations may occur during the descent process. We can also find the objective function $g_2\left(\mathbf{X}\right)$ converges in all three cases. In Fig.~\ref{obj convergence}, we also observe that the objective function value of $g_2\left(\mathbf{X}\right)$ initially decreases, then increases slightly, and finally converges. As $g_1\left(\mathbf{X}\right)$ decreases to a certain extent, it becomes apparent that a further decrease in $g_1\left(\mathbf{X}\right)$ leads to a slight increase in $g_2\left(\mathbf{X}\right)$. This phenomenon arises because $g_1\left(\mathbf{X}\right)$ and $g_2\left(\mathbf{X}\right)$ are to some extent mutually exclusive, indicating a tradeoff between optimizing covert performance and transmission performance.

\begin{figure}[!t]
\centering
\includegraphics[width=3.6in]{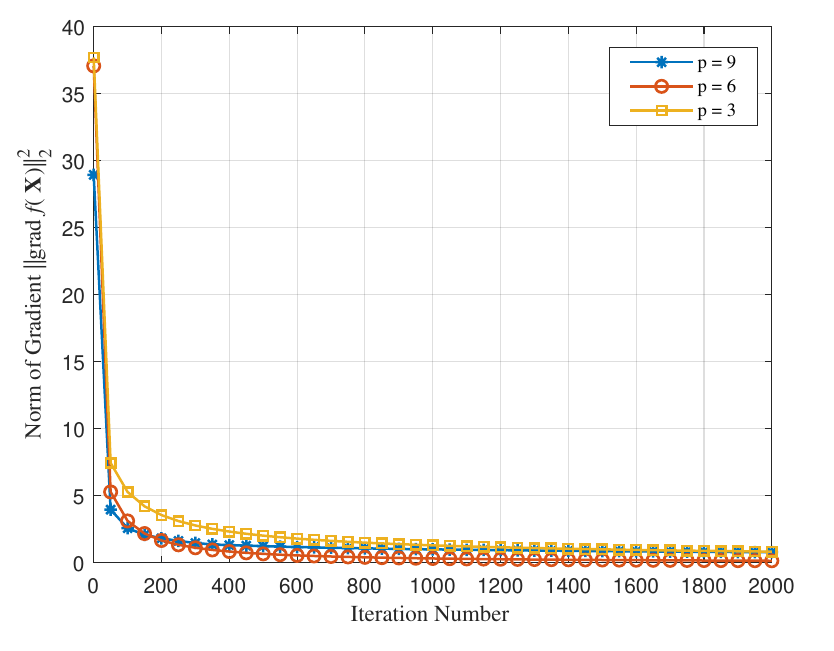}
\caption{$\|\operatorname{grad}f\left(\mathbf{X}\right)\|_2^2$ vs. iteration number.}
\label{gradient norm}
\end{figure}

Fig.~\ref{gradient norm} depicts the curves of the norm of the Riemannian gradient versus the iteration number for the R-SVRG algorithm under the above three different scenarios when $n = 18$. Since the corresponding $\mu$ for the problem differ under each scenario, the initial points are different. It can be observed from the Fig.~\ref{gradient norm} that in all three cases, the norm of the Riemannian gradient monotonically deceases and tends to zero. This numerical phenomenon validates {\it Theorem \ref{convergence theorem}} presented in this paper.

Combining Fig.~\ref{obj convergence} and Fig.~\ref{gradient norm}, we observe that the algorithm converges in fewer than 2000 iterations. With a summing term of $T=10000$, it indicates that we only need to compute the full gradient once at the initial point and fewer than 1000 gradient computations. In comparison to the computational burden of the cost gradient algorithm, our proposed R-SVRG algorithm requires less computation than two iterations of the full gradient, significantly reducing the computational overhead.

\begin{figure}[!t]
\centering
\includegraphics[width=3.6in]{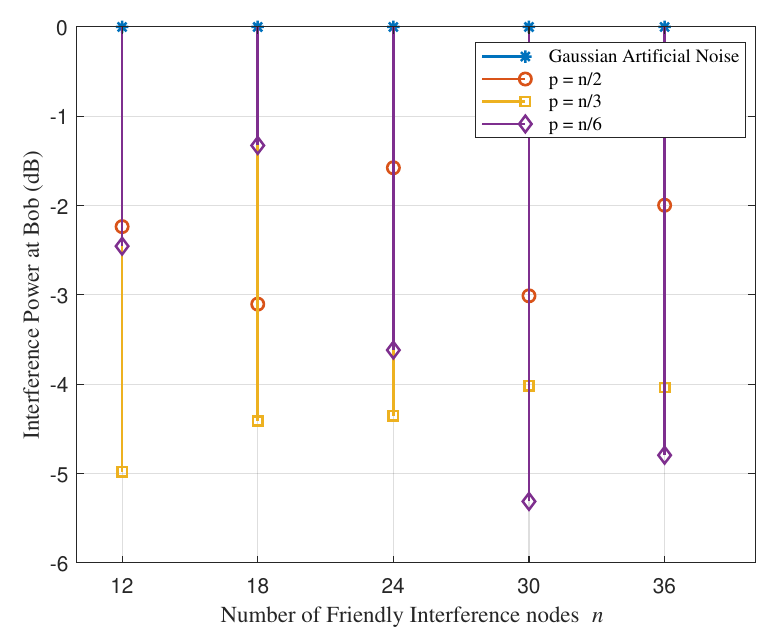}
\caption{Interference Power at Bob versus the number of friendly interference nodes $n$ for different $p$.}
\label{Interference Power}
\end{figure}

\begin{figure}[!t]
\centering
\includegraphics[width=3.7in]{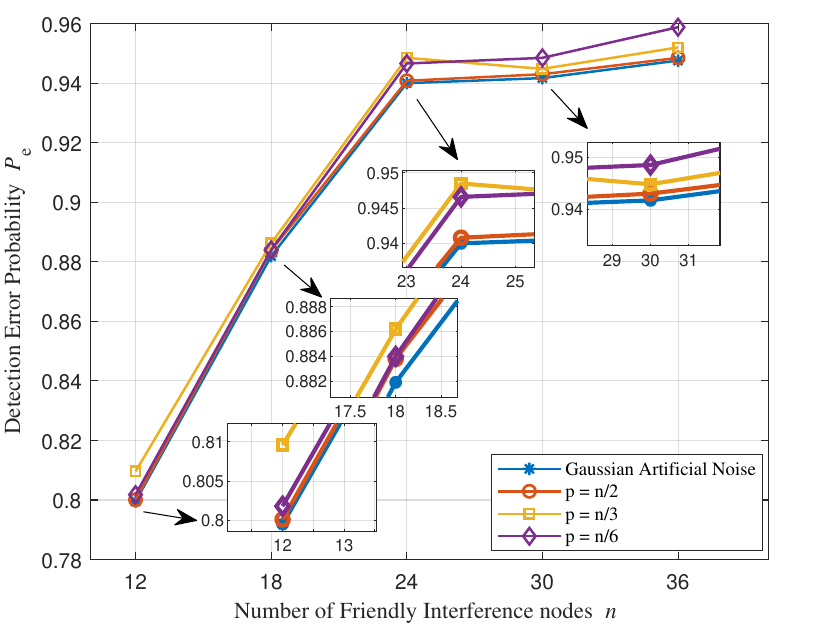}
\caption{$\mathbb{P}_{e}$ versus the number of friendly interference nodes $n$ for different $p$.}
\label{detection error}
\end{figure}

Fig.~\ref{Interference Power} and Fig.~\ref{detection error} respectively shows the interference power at Bob and the detection error probability $\mathbb{P}_{e}$ versus the number of friendly interference nodes $n$ for $p = n/2$, $p = n/3$ and $p = n/6$. Here the interference power at Bob is defined as
$$
P_i = 10 \log_{10}\frac{g_1\left(\mathbf{X}\right)}{g_1\left(\mathbf{I}_n\right)}
$$
where $g_1\left(\mathbf{X}\right)$ represents the interference power at Bob caused by our proposed artificial noise strategy, and $g_1\left(\mathbf{I}_n\right)$ represents the interference power at Bob caused by the Gaussian artificial noise strategy.

As seen from Fig.~\ref{Interference Power}, we have the following observations. Firstly, under the same number of friendly nodes, the performance of the artificial noise signals designed when $p = n/3$ is better than when $p = n/2$. This is because $p$ represents the rank of the basis matrix, and when $p$ is smaller, there are more optimization degrees of freedom, making it easier to find a basis matrix that minimizes the objective function. Secondly, when $n < 30$, the case where $p = n/6$ is not as good as when $p = n/3$, and even at $n = 18$, the performance of $p = n/6$ is worse than $p = n/2$ . This is because as $p$ decreases, corresponding to greater degrees of freedom for constraint $\mathbf{X}^H\mathbf{X} = \frac{n}{p}\mathbf{I}_p$, while $\frac{n}{p}$ also increases. When $n$ is small, the gain form decreasing $p$ cannot offset the increase in $\frac{n}{p}$.

From Fig.~\ref{detection error}, we also observe that when $n$ is small, the performance is optimal when $p = n/3$, and when n is large, the performance is optimal when $p = n/6$. Moreover, with the increase in the number of friendly interference nodes, the covert performance improves, as more interference nodes lead to greater interference, which is conducive to achieving covert communication.

Through Fig.~\ref{Interference Power} and Fig.~\ref{detection error}, we observe that with the artificial noise generation strategy proposed in this paper, the interference power at the Bob can reach around $-4$dB, ensuring comparable or even better covert performance than the Gaussian artificial noise strategy without optimization, regardless of the number of friendly interference nodes. This means that compared to Gaussian artificial noise strategy, the interference power at Bob is reduced by more than half. Numerical simulations demonstrate that our proposed strategy offers better covert performance.

\begin{figure}[!t]
\centering
\includegraphics[width=3.65in]{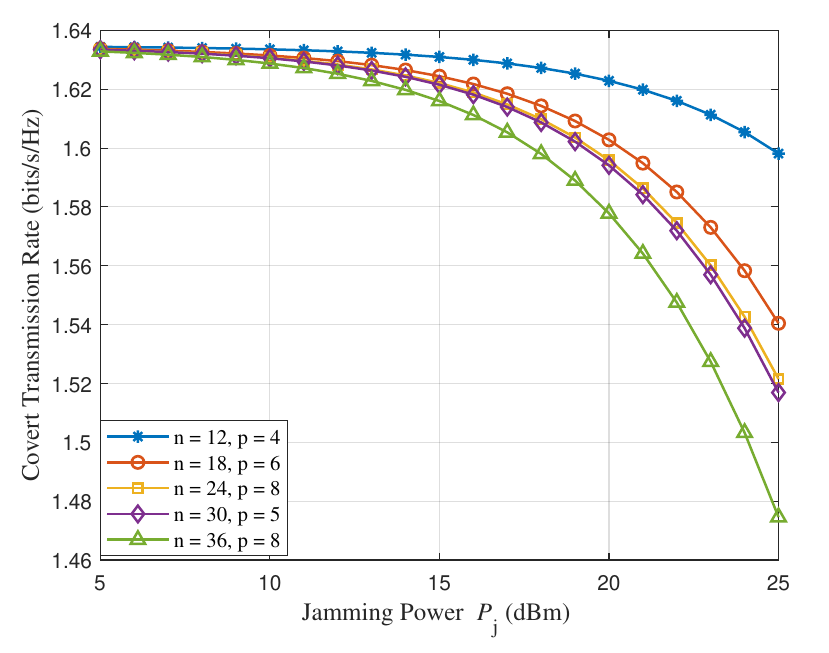}
\caption{Covert transmission rate versus the jamming power $P_j$ under different $n$ and $p$.}
\label{covert rate}
\end{figure}

Fig.~\ref{covert rate} demonstrates the covert transmission rate under different $n$ and $p$ versus the jamming power $P_j$. Here we choose to plot the optimal performance for different $n$ values. It can be observed that the higher the jamming power, the lower the covert transmission rate, and as the jamming power increases, the transmission rate decreases more rapidly. Moreover, as $n$ increases, the covert transmission rate decreases.

\begin{figure}[!t]
\centering
\includegraphics[width=3.6in]{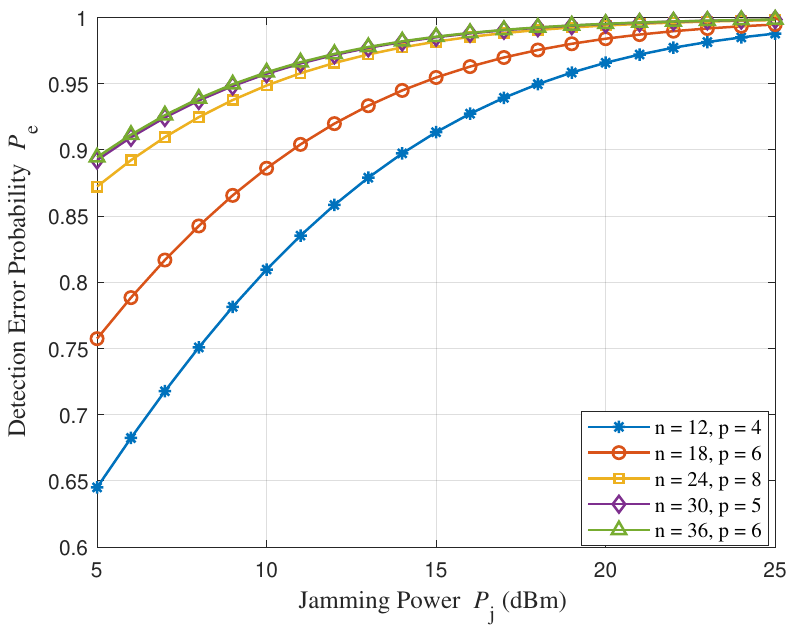}
\caption{$\mathbb{P}_{e}$ versus the jamming power $P_j$ under different $n$ and $p$.}
\label{covert performance}
\end{figure}

Fig.~\ref{covert performance} demonstrates the detection error probability $\mathbb{P}_{e}$ under different $n$ and $p$ versus the jamming power $P_j$. As observed from this figure, we find that the larger the jamming power, the greater the detection error probability, indicating higher covertness. Additionally, as the jamming power increases, the growth rate of the detection error probability slows down. Furthermore, as $n$ increases, the level of covert performance improves.

\section{CONCLISION}
In this paper, we investigate a scenario of covert communication assisted by multiple friendly interference nodes. We design artificial noise using imperfect CSI to significantly degrade Willie's detection performance while minimizing its impact on legitimate communication, thereby enhancing the system's covert performance. To handle the non-convex unitary matrix constraints, we adopt a Riemannian manifold optimization framework, leveraging the geometric structure of constraints to embed them into the search space and solve the problem as an unconstrained one. Additionally, we employ a stochastic gradient algorithm to reduce the computational cost per iteration. The proposed algorithm's convergence is theoretically proven. Numerical simulations validate a significant improvement in covert performance with the proposed artificial noise strategy.
\appendices
\section{} \label{lambda proof}
Under the null hypothesis $\mathcal{H}_0$
\begin{equation}\label{lambda 0}
\begin{aligned}
\lambda_0 & = \mathbb{E}\left[\left(\sqrt{P_j}\mathbf{h}_{j,w}^H\mathbf{v}_j+n_w\right)^2\right]\\
& = \mathbb{E}\left(P_j\mathbf{h}_{j,w}^H\mathbf{v}_j\mathbf{v}_j^H\mathbf{h}_{j,w}\right)+\sigma_w^2\\
& = \mathbb{E}\left(P_j\mathbf{h}_{j,w}^H\mathbf{X}\mathbf{a}\mathbf{a}^H\mathbf{X}^H\mathbf{h}_{j,w}\right)+\sigma_w^2\\
& = P_j \mathbf{h}_{j, w}{ }^H \mathbf{X} \mathbf{X}^H \mathbf{h}_{j, w}+\sigma_w^2
\end{aligned}
\end{equation}

Under the alternative hypothesis $\mathcal{H}_1$
\begin{equation}\label{lambda 1}
\begin{aligned}
\lambda_1 & = \mathbb{E}\left[\left(\sqrt{P_a}h_{a,w}s+\sqrt{P_j}\mathbf{h}_{j,w}^H\mathbf{v}_j+n_w\right)^2\right]\\
& = P_a|h_{a,w}|^2+\mathbb{E}\left(P_j\mathbf{h}_{j,w}^H\mathbf{v}_j\mathbf{v}_j^H\mathbf{h}_{j,w}\right)+\sigma_w^2\\
& =P_a|h_{a,w}|^2 + P_j \mathbf{h}_{j, w}{ }^H \mathbf{X} \mathbf{X}^H \mathbf{h}_{j, w}+\sigma_w^2
\end{aligned}
\end{equation}

From this, we obtain \eqref{H0 variance}, \eqref{H1 variance}. $\hfill\blacksquare$

\section{PROOF OF FACT \ref{fact-complex-real}}\label{fact1 proof}
For any complex matrix $\mathbf{X}\in\mathbb{C}^{n \times p}$, we can represent it using a real matrix $\bar{\mathbf{X}}$ given in \eqref{complex matrix real expression}.
It is easy to proof that $\bar{\mathbf{X}}$ possesses property $\bar{\mathbf{X}}\mathbf{J}_p = \mathbf{J}_n\mathbf{X}$, thus $\bar{\mathbf{X}} \in \mathcal{S P}\left(p,n\right)$.

From \eqref{stp manifold}, the real representation of $\operatorname{St}\left(p,n,\mathbb{C}\right)$ is $\operatorname{Stp}\left(p,n\right) =: \operatorname{St}\left(2p,2n,\mathbb{R}\right) \cap \mathcal{S P}\left(p,n\right)$. We can easily obtain that $\bar{\mathbf{X}} \in \operatorname{Stp}\left(p,n\right)$ is equivalent to $\bar{\mathbf{X}} \in \operatorname{St}\left(2p,2n,\mathbb{R}\right)$, which means $\mathbf{X} \in \operatorname{St}\left(p,n,\mathbb{C}\right)$ is equivalent to $\bar{\mathbf{X}} \in \operatorname{St}\left(2p,2n,\mathbb{R}\right)$ and vice versa. This concludes the proof. $\hfill\blacksquare$

\section{PROOF OF TANGENT SPACE, RIEMANNIAN METRIC AND THE ORTHOGONAL COMPLEMENT SPACE}\label{complex stiefel proof}
Based on the real representation of $\operatorname{St}\left(p,n,\mathbb{C}\right)$ is $\operatorname{Stp}\left(p,n\right)$, we can use the geometric properties of $\operatorname{Stp}\left(p,n\right)$ to characterize those of $\operatorname{St}\left(p,n,\mathbb{C}\right)$. Following are the proofs regarding the tangent space, Riemannian metric and the orthogonal complement space of $\operatorname{Stp}\left(p,n\right)$.

The tangent space to $\operatorname{St}\left(2p.2n,\mathbb{R}\right)$ at $\mathbf{Y}$ is \cite{manifold-optimization-absil}
\begin{equation}\label{tangent space real stiefel}
T_{\mathbf{Y}}\operatorname{St}\left(2p,2n,\mathbb{R}\right) = \{\xi_{\mathbf{Y}} \in \mathbb{R}^{2n \times 2p}:\mathbf{Y}^T\xi_{\mathbf{Y}}+\xi_{\mathbf{Y}}^T\mathbf{Y} = 0\}
\end{equation}

So we have
\begin{equation}\label{tangent space Stp}
 T_{\bar{\mathbf{X}}}\operatorname{Stp}(p, n)
= \{\bar{\xi}_{\bar{\mathbf{X}}} \in \mathcal{S} \mathcal{P}(p, n) \mid \left.\bar{\xi}_{\bar{\mathbf{X}}}^T \bar{\mathbf{X}}+\bar{\mathbf{X}}^T \bar{\xi}_{\bar{\mathbf{X}}}=0\right\}
\end{equation}

The inner product of the embedding space $\mathbb{R}^{2n \times 2p}$ is
\begin{equation}\label{Euclidean inner product}
\langle\mathbf{Y}_1,\mathbf{Y}_2\rangle = \operatorname{tr}\left(\mathbf{Y}_1^T\mathbf{Y}_2\right)
\end{equation}
where $\mathbf{Y}_1, \mathbf{Y}_2 \in \mathbb{R}^{2n \times 2p}$.

When restricted to the subspace $\operatorname{Stp}\left(p,n\right)$ of $\mathbb{R}^{2n \times 2p}$, the inner product has the form
\begin{equation}\label{embedding inner product}
\begin{aligned}
\langle\bar{\mathbf{X}}_1, \bar{\mathbf{X}}_2\rangle &=\operatorname{tr}\left(\bar{\mathbf{X}}_1^T\bar{\mathbf{X}}_2\right)\\
&=2\operatorname{tr}\left(\operatorname{Re}\left(\mathbf{X}_1\right)^T
\operatorname{Re}\left(\mathbf{X}_2\right)+\operatorname{Im}\left(\mathbf{X}_1\right)^T
\operatorname{Im}\left(\mathbf{X}_2\right)\right)
\end{aligned}
\end{equation}
for $\bar{\mathbf{X}}_1=\left(\begin{array}{cc}\operatorname{Re}\left(\mathbf{X}_1\right) & \operatorname{Im}\left(\mathbf{X}_1\right)  \\ -\operatorname{Im}\left(\mathbf{X}_1\right)  & \operatorname{Re}\left(\mathbf{X}_1\right) \end{array}\right)$ and $ \bar{\mathbf{X}}_2=\left(\begin{array}{cc}\operatorname{Re}\left(\mathbf{X}_2\right) & \operatorname{Im}\left(\mathbf{X}_2\right)  \\ -\operatorname{Im}\left(\mathbf{X}_2\right)  & \operatorname{Re}\left(\mathbf{X}_2\right) \end{array}\right)$.

Getting rid of the factor 2 in the right-hand side of \eqref{embedding inner product}, we define the inner product on $\mathcal{S P}(p, n)$ to be
\begin{equation}
\langle\bar{\mathbf{X}}_1, \bar{\mathbf{X}}_2\rangle =\frac{1}{2}\operatorname{tr}\left(\bar{\mathbf{X}}_1^T\bar{\mathbf{X}}_2\right)
\end{equation}

It is easy to get that the Riemannian metric on the tangent space is given by \eqref{Riemannian metric}.

The orthogonal complement space of the tangent space has the property
\begin{equation}\label{orthogonal property}
g_{\bar{\mathbf{x}}}(\bar{\xi}_{\bar{\mathbf{X}}}, \bar{\zeta}_{\bar{\mathbf{X}}})=\frac{1}{2}\operatorname{tr}\left(\bar{\xi}_{\bar{\mathbf{X}}}^T \bar{\zeta}_{\bar{\mathbf{X}}}\right)=0
\end{equation}
where $\bar{\xi}_{\bar{\mathbf{X}}} \in T_{\bar{\mathbf{X}}} \operatorname{Stp}(p, n)$, $\bar{\zeta}_{\bar{\mathbf{X}}} \in \left(T_{\bar{\mathbf{X}}} \operatorname{Stp}(p, n)\right)^{\perp}$.

By obserbing \eqref{tangent space Stp}, \eqref{orthogonal property}, we set the element of the orthogonal complement of the tangent space $T_{\bar{\mathbf{X}}}(\operatorname{Stp}(p, n))$ is $\bar{\mathbf{X}}\mathbf{S}$, we have
\begin{equation}
\operatorname{tr}\left(\bar{\xi}_{\bar{\mathbf{X}}}^T \bar{\mathbf{X}}\mathbf{S}\right) = 0
\end{equation}

Then, we have:
\begin{equation}
\operatorname{tr}\left(\bar{\xi}_{\bar{\mathbf{X}}}^T \bar{\mathbf{X}}\mathbf{S}+\mathbf{S}^T\bar{\mathbf{X}}^T\bar{\xi}_{\bar{\mathbf{X}}}\right) = 0
\end{equation}

From \eqref{tangent space Stp}, we can get $\mathbf{S}^T = \mathbf{S}$. Therefore, the orthogonal complement space of the tangent space is \eqref{Stp orthogonal complement space}. This concludes the proof. $\hfill\blacksquare$

\section{PROOF OF LEMMA \ref{lemma-Stp-gradient}}\label{lemma1 proof}
From reference \cite{manifold-optimization-absil}, it is known that the orthogonal complement space of the tangent space of $\operatorname{St}\left(2p,2n,\mathbb{R}\right)$ at $\mathbf{Y}$ is
\begin{equation}\label{real stiefel orthogonal}
\left(T_{\mathbf{Y}}\operatorname{St}\left(2p,2n,\mathbb{R}\right)\right)^{\perp}=\{\mathbf{Y}\mathbf{S}: S\in\mathcal{S}_{sym}\left(2p\right)\}
\end{equation}
where $S_{s y m}(2p)$ is a set of $2 p \times 2 p$ dimensional real symmetric matrices.

Observing the form of the tangent space and the orthogonal complement space of $\operatorname{Stp}\left(p,n\right)$, we can find similarities with the geometric properties of $\operatorname{St}\left(p,n,\mathbb{R}\right)$.

The difference compared to \eqref{tangent space real stiefel}, \eqref{tangent space Stp} is that it needs to satisfy $\bar{\xi}_{\bar{\mathbf{X}}} \in \mathcal{SP}\left(p,n\right)$, and the difference compared to \eqref{Stp orthogonal complement space}, \eqref{real stiefel orthogonal} is that it satisfies $\bar{\mathbf{X}} \in \mathcal{SP}\left(p,n\right)$. This property is inherently possessed by $\bar{\xi}_{\bar{\mathbf{X}}}, \bar{\mathbf{X}}$.

For manifold $\operatorname{St}\left(2p,2n,\mathbb{R}\right)$, the Riemannian gradient is given by
\begin{equation}\label{real stiefel rgrad}
\operatorname{grad} f(\mathbf{Y})=\left(\mathbf{I}-\mathbf{Y} \mathbf{Y}^T\right) \nabla f(\mathbf{Y})+\mathbf{Y}  \text{skew}\left(\mathbf{Y}^T \nabla f(\mathbf{Y})\right)
\end{equation}

Similar to $\operatorname{St}\left(2p,2n,\mathbb{R}\right)$, we can derive the Riemannian gradient of manifold $\operatorname{Stp}\left(p,n\right)$ as \eqref{Stp rgrad}, and it can be easily proven that $\operatorname{grad} \bar{f}(\bar{\mathbf{X}}) \in  T_{\bar{\mathbf{X}}}\operatorname{Stp}(p, n)$. This concludes the proof. $\hfill\blacksquare$

\section{PROOF OF LEMMA \ref{lemma-complex-stiefel-gradient}}\label{lemma2 proof}
We can easily obtain:
\begin{equation}\label{Euclidean gradient real expression}
\nabla\bar{f}\left(\bar{\mathbf{X}}\right)=\left(\begin{array}{cc}\operatorname{Re}\left(\nabla f\left(\mathbf{X}\right)\right) & \operatorname{Im}\left(\nabla f\left(\mathbf{X}\right)\right) \\ -\operatorname{Im}\left(\nabla f\left(\mathbf{X}\right)\right) & \operatorname{Re}\left(\nabla f\left(\mathbf{X}\right)\right)\end{array}\right)
\end{equation}
Substituting \eqref{complex matrix real expression}, \eqref{Euclidean gradient real expression} into \eqref{Stp rgrad}, we get
\begin{equation}\label{Stp rgrad part1}
\begin{aligned}
&\left(\mathbf{I}-\bar{\mathbf{X}}\bar{\mathbf{X}}^T\right)\nabla\bar{f}\left(\bar{\mathbf{X}}\right) = \\
&\left(\!\!\begin{array}{cc}\operatorname{Re}\left(\left(\mathbf{I}-\mathbf{X}\mathbf{X}^H\right)\nabla f\left(\mathbf{X}\right)\right) & \operatorname{Im}\left(\left(\mathbf{I}-\mathbf{X}\mathbf{X}^H\right)\nabla f\left(\mathbf{X}\right)\right) \\ -\operatorname{Im}\left(\left(\mathbf{I}-\mathbf{X}\mathbf{X}^H\right)\nabla f\left(\mathbf{X}\right)\right) &\operatorname{Re}\left(\left(\mathbf{I}-\mathbf{X}\mathbf{X}^H\right)\nabla f\left(\mathbf{X}\right)\right)\end{array}\!\!\right)
\end{aligned}
\end{equation}

\begin{equation}\label{Stp rgrad part2}
\begin{aligned}
&\bar{\mathbf{X}}\operatorname{skew}\left(\bar{\mathbf{X}}^T\nabla \bar{f}\left(\bar{\mathbf{X}}\right)\right) = \\
&\left(\!\!\!\begin{array}{cc}\operatorname{Re}\left(\mathbf{X}\operatorname{skew}\left(\mathbf{X}^H\nabla f\left(\mathbf{X}\right)\right)\right) \! & \! \operatorname{Im}\left(\mathbf{X}\operatorname{skew}\left(\mathbf{X}^H\nabla f\left(\mathbf{X}\right)\right)\right) \\ -\operatorname{Im}\left(\mathbf{X}\operatorname{skew}\left(\mathbf{X}^H\nabla f\left(\mathbf{X}\right)\right)\right) \! & \!\operatorname{Re}\left(\mathbf{X}\operatorname{skew}\left(\mathbf{X}^H\nabla f\left(\mathbf{X}\right)\right)\right)\end{array}\!\!\!\right)
\end{aligned}
\end{equation}

Therefore, we can get
\begin{equation}\label{rgrad complex to real}
\operatorname{grad} \bar{f}(\bar{\mathbf{X}})= \left(\begin{array}{cc}\operatorname{Re}\left(\operatorname{grad} f(\mathbf{X})\right) & \operatorname{Im}\left(\operatorname{grad} f(\mathbf{X})\right) \\ -\operatorname{Im}\left(\operatorname{grad} f(\mathbf{X})\right) & \operatorname{Re}\left(\operatorname{grad} f(\mathbf{X})\right)\end{array}\right)
\end{equation}

This concludes the proof. $\hfill\blacksquare$

\section{PROOF OF LEMMA \ref{lemma-Stp-retraction}}\label{lemma3 proof}
On the real Stiefel manifold $\operatorname{St}\left(2p,2n,\mathbb{R}\right)$, there exists a retraction based on the QR decomposition, which is defined to be
\begin{equation}\label{real stiefel retraction}
R_{\mathbf{Y}}\left(\xi_{\mathbf{Y}}\right) = \operatorname{qf}\left(\mathbf{Y}+\xi_{\mathbf{Y}}\right)
\end{equation}
where $\mathbf{Y} \in \operatorname{St}\left(2p,2n,\mathbb{R}\right)$, $\xi_{\mathbf{Y}} \in T_{\mathbf{Y}}\operatorname{St}\left(2p,2n,\mathbb{R}\right)$ and where $\operatorname{qf}\left(\cdot\right)$ denotes the Q-factor of the QR decomposition of the matrix.

However \eqref{real stiefel retraction} cannot apply for $\operatorname{Stp}\left(p,n\right)$. That is because $\bar{\xi}_{\bar{\mathbf{X}}} \in \mathcal{SP}\left(p,n\right)$, $\operatorname{qf}\left(\bar{\xi}_{\bar{\mathbf{X}}}\right)$ no longer belongs to $\mathcal{SP}\left(p,n\right)$.

Next, we observe that performing a complex QR decomposition on the complex representation of the real matrix $\bar{\mathbf{X}}+\bar{\xi}_{\bar{\mathbf{X}}}$ yields
\begin{equation}\label{complex QR}
\bar{\mathbf{X}}+\bar{\xi}_{\bar{\mathbf{X}}} = \mathbf{QR}
\end{equation}
where $\mathbf{Q} \in \operatorname{St}\left(p,n,\mathbb{C}\right), \mathbf{R} \in S_{upp}^{+}\left(p\right)$ and $S_{upp}^{+}\left(p\right)$ denotes the set of all $p \times p$ upper triangular matrices with strictly positive diagonal entries, then $\operatorname{qf}\left(\bar{\mathbf{X}}+\bar{\xi}_{\bar{\mathbf{X}}}\right) = \mathbf{Q}$.

We then define the QR-based retraction on $\operatorname{Stp}\left(p,n\right)$ as \eqref{Stp retraction} and \eqref{Stp retraction} satisfies the definition of retraction given in reference \cite{manifold-optimization-absil}. This concludes the proof. $\hfill\blacksquare$

\section{PROOF OF THEOREM \ref{convergence theorem}}\label{convergence proof}
The sequence genrated by the proposed R-SVRG algorithm, denotes as $\{\mathbf{X}_q^l\}$ exists in a set $K \subset \operatorname{St}\left(p,n,\mathbb{C}\right)$ such that for all $l,q \geq 0$, we have $\mathbf{X}_q^l \in K$.

Here, we first prove that the set $K$ is a connected and compact topological space. Then, we demonstrate the convergence of $\{f\left(\mathbf{X}_q^l\right)\}$, following by proving $\displaystyle\lim _{l \rightarrow \infty} \alpha_q^l\left\|\operatorname{grad} f\left(\mathbf{X}_q^l\right)\right\|_{\mathbf{x}_q^l}^2=0$. Finally, we establish the convergence of $\|\operatorname{grad} f\left(\mathbf{X}_q^l\right)\|_{\mathbf{X}_q^l}^2$ and $\displaystyle\lim _{l \rightarrow \infty} \left\|\operatorname{grad} f\left(\mathbf{X}_q^l\right)\right\|_{\mathbf{X}_q^l}^2=0$.

Since the search space of this algorithm is not the familiar Euclidean space, we need to prove the concept of convergence in the constructed Riemannian manifold, i.e., in this set there exists the intermediate value theorem, the extreme value theorem, and the uniform continuity theorem. According to \cite{topology}, the intermediate value theorem depends on the connectedness of the topological space, and the extreme value theorem and the uniform continuity theorem depend on the compactness of the topological space. Here, we prove that the topological space of the set $K$ is a connected and compact topological space.

{\it Connectedness:} A partition of the topological space of set $K$ is defined as a pair of sets whose union equals $K$ and are disjoint. It is easy to see that there is no partition of the set $K$, hence the topological space of set $K$ is a connected space.

{\it Compactness:}  By analyzing the set $K$, we can first observe that this set is a closed set, since the real representation of the complex manifold $\operatorname{St}\left(p,n,\mathbb{C}\right)$ is  $\operatorname{Stp}\left(p,n\right):=\operatorname{St}\left(2p,2n,\mathbb{R}\right) \cap \mathcal{SP}\left(p,n\right)$, where $\mathbf{X} \in \operatorname{Stp}\left(p,n\right)$, it must be that $\mathbf{X} \in \operatorname{St}\left(2p,2n,\mathbb{R}\right)$. Moreover, $\operatorname{St}\left(2p,2n,\mathbb{R}\right)$ is bounded with respect to the Euclidean metric, and since $K \subset \operatorname{St}(p, n,\mathbb{C})$, the set $K$ is also bounded with respect to the Euclidean metric. Therefore, any open cover of $K$ contains a finite subfamily that covers $K$. Thus, the space of set $K$ is a compact space.

Here, we have shown that the set $K$ formed by the iterative variables satisfies the prerequisite conditions for proving convergence.

Next, we prove the existence of $\tilde{\eta}_{q-1}^l \in T_{\tilde{\mathbf{X}}^{l-1}}\operatorname{St}\left(p,n,\mathbb{C}\right)$ connecting points $\mathbf{X}_{q-1}^l$ and $\tilde{\mathbf{X}}^{l-1}$, such that $R_{\tilde{\mathbf{X}}^{l-1}}\left(\tilde{\eta}_{q-1}^l\right) = \mathbf{X}_{q-1}^l$.

According to \eqref{complex stiefel retraction}, we have $\mathbf{X}_{q-1}^l = \operatorname{qf}\left(\tilde{\mathbf{X}}^{l-1}+\tilde{\eta}_{q-1}^l\right)$, implying
\begin{equation}\label{QR equation}
\tilde{\mathbf{X}}^{l-1}+\tilde{\eta}_{q-1}^l=\mathbf{X}_{q-1}^l \mathbf{R}
\end{equation}
where $\mathbf{R}$ is an upper triangle matrix with strictly positive diagonal elements.

Given any matrices $\mathbf{X}_{q-1}^l, \widetilde{\mathbf{X}}^{l-1} \in \operatorname{St}\left(p, n,\mathbb{C}\right)$, the inverse map of contracting $R_{\widetilde{\mathbf{X}}^{l-1}}\left(\cdot\right)$ is
\begin{equation}\label{inverse map of R}
R_{\widetilde{\mathbf{X}}^{l-1}}^{-1}\left(\mathbf{X}_{q-1}^l\right)=\mathbf{X}_{q-1}^l \mathbf{R}-\widetilde{\mathbf{X}}^{l-1}
\end{equation}

Left multiplying \eqref{QR equation} by $\left(\widetilde{\mathbf{X}}^{l-1}\right)^H$ yields
\begin{equation}\label{equation after multiple}
\mathbf{I}+\left(\widetilde{\mathbf{X}}^{l-1}\right)^H \tilde{\eta}_{q-1}^l=\left(\widetilde{\mathbf{X}}^{l-1}\right)^H \mathbf{X}_{q-1}^l \mathbf{R}
\end{equation}

Transposing both sides of the equation conjointly
\begin{equation}\label{equation after transpose}
\mathbf{I}+\left(\tilde{\eta}_{q-1}^l\right)^H \widetilde{\mathbf{X}}^{l-1}=\mathbf{R}^H\left(\mathbf{X}_{q-1}^l\right)^H \widetilde{\mathbf{X}}^{l-1}
\end{equation}

Adding \eqref{inverse map of R} and \eqref{equation after multiple} together yields
\begin{equation}\label{equation after add}
\begin{aligned}
&2 \mathbf{I}+\left(\widetilde{\mathbf{X}}^{l-1}\right)^H \tilde{\eta}_{q-1}^l+\left(\tilde{\eta}_{q-1}^l\right)^H \widetilde{\mathbf{X}}^{l-1}\\
&=\mathbf{R}^H\left(\mathbf{X}_{q-1}^l\right)^H \widetilde{\mathbf{X}}^{l-1}+\left(\widetilde{\mathbf{X}}^{l-1}\right)^H \mathbf{X}_{q-1}^l \mathbf{R}
\end{aligned}
\end{equation}

According to \eqref{complex stiefel rgrad}
\begin{equation}\label{equation by rgrad}
\begin{aligned}
&\tilde{\eta}_{q-1}^l=\left(\mathbf{I}-\tilde{\mathbf{X}}^{l-1}\left(\tilde{\mathbf{X}}^{l-1}\right)^H\right) \eta_{q-1}^l+\\
&\frac{1}{2} \tilde{\mathbf{X}}^{l-1}\left(\left(\tilde{\mathbf{X}}^{l-1}\right)^H \eta_{q-1}^l-\left(\left(\tilde{\mathbf{X}}^{l-1}\right)^H \eta_{q-1}^l\right)^H\right)
\end{aligned}
\end{equation}
where: $\eta_{q-1}^l$ is the Euclidean gradient corresponding to the Riemannian gradient $\tilde{\eta}_{q-1}^l$ and
\begin{equation}\label{skew-symmetric part}
\begin{aligned}
\left(\tilde{\mathbf{X}}^{l-1}\right)^H \tilde{\eta}_{q-1}^l & =\frac{1}{2}\left(\left(\tilde{\mathbf{X}}^{l-1}\right)^H \eta_{q-1}^l-\left(\left(\tilde{\mathbf{X}}^{l-1}\right)^H \eta_{q-1}^l\right)^H\right) \\
& =\operatorname{skew}\left(\left(\tilde{\mathbf{X}}^{l-1}\right)^H \eta_{q-1}^l\right)
\end{aligned}
\end{equation}
is skew-symmetric matrix.

Therefore,
\begin{equation}\label{zero part}
\left(\tilde{\mathbf{X}}^{l-1}\right)^H \tilde{\eta}_{q-1}^l+\left(\tilde{\eta}_{q-1}^l\right)^H \widetilde{\mathbf{X}}^{l-1}=0
\end{equation}

\eqref{equation after add} becomes
\begin{equation}\label{final equation}
\mathbf{R}^H\left(\mathbf{X}_{q-1}^l\right)^H \widetilde{\mathbf{X}}^{l-1}+\left(\widetilde{\mathbf{X}}^{l-1}\right)^H \mathbf{X}_{q-1}^l \mathbf{R}=2 \mathbf{I}
\end{equation}

The solution for $\mathbf{R}$ can be obtained through \cite{stiefel-manifold-tsp}, and finally, based on \eqref{QR equation}, we can obtain $\tilde{\eta}_{q-1}^l$. Therefore, $\tilde{\eta}_{q-1}^l$ exists.

Since the space of set $K$ is compact, any continuous function on $K$ is bounded. Therefore, there exists
\begin{equation}\label{rgrad bound sample}
\left\|\operatorname{grad} f_{t_q^l}\left(\mathbf{X}_q^l\right)\right\|_{\mathbf{X}_q^l} \leq C_1
\end{equation}
\begin{equation}\label{rgrad bound vectrans sample}
\left\|\mathcal{T}_{\tilde{\eta}_{q-1}^l}\left(\operatorname{grad} t_{t_q^l}\left(\widetilde{\mathbf{X}}^{l-1}\right)\right)\right\|_{\mathbf{X}_{q-1}^l} \leq C_2
\end{equation}
\begin{equation}\label{rgrad bound vectrans full}
\left\|\mathcal{T}_{\tilde{\eta}_{q-1}^l}\left(\operatorname{grad} f\left(\widetilde{\mathbf{X}}^{l-1}\right)\right)\right\|_{\mathbf{X}_{q-1}^l} \leq C_3
\end{equation}

According to the triangle inequality, we have
\begin{equation}\label{triangle inequality R-SVRG gradient bound}
\begin{aligned}
&\left\|\xi_{q+1}^l\right\|_{\mathbf{X}_q^l} =\\
&\left\|\operatorname{grad}\! f_{t_q^l}\left(\mathbf{X}_q^l\right)\!-\!\mathcal{T}_{\tilde{\eta}_{q-1}^l}\!\!\left(\!\operatorname{grad} f_{t_q^l}\left(\widetilde{\mathbf{X}}^{l-1}\!\right)\!-\!\operatorname{grad} f\left(\widetilde{\mathbf{X}}^{l-1}\right)\!\right)\!\right\|_{\mathbf{X}_q^l} \\
& \leq C_1+C_2+C_3 =C^{\prime}
\end{aligned}
\end{equation}

According to \eqref{stepsize condition 1}, \eqref{stepsize condition 2}, we can obtain that the maximum step size is $\alpha_0^1$. Hence, there exists $I > 0$ such that $\alpha_q^l C^{\prime} < \alpha_0^1 C^{\prime} < I$, ensuring that for any $\mathbf{X}_q^l \in K$, $R_{\mathbf{X}_q^l} (\cdot)$ defined at an origin $\mathbf{0}_{\mathbf{X}_q^l}$ in $T_{\mathbf{X}_q^l} \mathrm{St}(p, n,\mathbb{C})$, with a radius of $I$, lies within the ball $\mathbb{B}(\mathbf{0}_{\mathbf{X}_q^l}, I) \subset T_{\mathbf{X}_q^l} \mathrm{St}(p, n,\mathbb{C})$.

Define the function $g\left(\tau ; \mathbf{X}_q^l, \xi_{q+1}^l\right)=f \circ R_{\mathbf{X}_q^l}\left(-\tau \xi_{q+1}^l\right)$. According to \cite{manifold-optimization-absil}, the retraction based on the $\mathrm{QR}$ decomposition can be obtained through the Gram-Schmidt process, thus it is $C^{\infty}$. Hence, there exists a constant $k_1$ such that
\begin{equation}\label{smooth}
\frac{d^2}{d \tau^2} g\left(\tau ; \mathbf{X}_q^l, \xi_{q+1}^l\right) \leq 2 k_1
\end{equation}
where $\tau \in\left[0, \alpha_0^1\right], \xi_{q+1}^l \in \mathbb{B}\left(\mathbf{0}_{\mathbf{x}_q^l}, I\right), \mathbf{X}_q^l \in K$

Since set $\left[0, \alpha_0^1\right] \! \times \! \left\{\!\left(\mathbf{X}_q^l, \xi_{q+1}^l\right) \! \mid \! \xi_{q+1}^l \in \mathbb{B}\left(\mathbf{0}_{\mathbf{X}_q^l}, I\right), \mathbf{X}_q^l \in K \!\right\}$ is compact, according to the Taylor expansion of integral remainder, we have
\begin{equation}\label{Taylor expansion of integral remainder}
\begin{aligned}
&g\left(\alpha_q^l ; \mathbf{X}_q^l, \xi_{q+1}^l\right)= g\left(0 ; \mathbf{X}_q^l, \xi_{q+1}^l\right)+g^{\prime}\left(0 ; \mathbf{X}_q^l, \xi_{q+1}^l\right) \alpha_q^l+ \\
&\int_0^{\alpha_q^l} g^{\prime \prime}\left(\tau ; \mathbf{X}_q^l, \xi_{q+1}^l\right)\left(\alpha_q^l-\tau\right) d \tau
\end{aligned}
\end{equation}

Let $\tau_1 = \frac{\tau}{\alpha_q^l}$, then \eqref{Taylor expansion of integral remainder} becomes
\begin{equation}\label{Taylor expansion reformulation}
\begin{aligned}
&g\left(\alpha_q^l ; \mathbf{X}_q^l, \xi_{q+1}^l\right)=g\left(0 ; \mathbf{X}_q^l, \xi_{q+1}^l\right)+g^{\prime}\left(0 ; \mathbf{X}_q^l, \xi_{q+1}^l\right) \alpha_q^l+\\
&\left(\alpha_q^l\right)^2 \int_0^1\left(1-\tau_1\right) g^{\prime \prime}\left(\alpha_q^l \tau_1 ; \mathbf{X}_q^l, \xi_{q+1}^l\right) d \tau_1
\end{aligned}
\end{equation}

From \eqref{Taylor expansion reformulation}, we have
\begin{equation}\label{fX decrease}
\begin{aligned}
& f\left(\mathbf{X}_{q+1}^l\right)-f\left(\mathbf{X}_q^l\right) \\
& =f\left(R_{\mathbf{X}_{q-1}^l}\left(-\alpha_q^l \xi_{q+1}^l\right)\right)-f\left(\mathbf{X}_q^l\right) \\
& =g\left(\alpha_q^l ; \mathbf{X}_q^l, \xi_{q+1}^l\right)-g\left(0 ; \mathbf{X}_q^l, \xi_{q+1}^l\right) \\
& =-\alpha_q^l\!\left\langle\xi_{q+1}^l, \operatorname{grad}\! f\left(\mathbf{X}_q^l\right)\!\right\rangle_{\mathbf{X}_q^l}\!\!+\!\left(\alpha_q^l\right)^2 \!\!\int_0^1\!\!\!\left(1-\tau_1\right) g^{\prime \prime}\!\left(\alpha_q^l \tau_1\right) d \tau_1 \\
& \leq-\alpha_q^l\left\langle\xi_{q+1}^l, \operatorname{grad} f\left(\mathbf{X}_q^l\right)\right\rangle_{\mathbf{X}_q^l}+\left(\alpha_q^l\right)^2 k_1
\end{aligned}
\end{equation}

Since each iteration involves randomly selecting sample gradients, $f\left(\mathbf{X}_q^l\right)$ is a stochastic process. Next, we derive conclusions about the convergence of the function $f\left(\mathbf{X}_q^l\right)$ by analyzing some properties of this stochastic process.

Constructing the conditional expectation $\mathbb{E}\left[\xi_{q+1}^l \mid \mathcal{F}_{q+1}^l\right]$, we aim to have all the information about the values of random variables from previous iterations at the current iteration given the condition $\mathcal{F}_{q+1}^l$.

Here, $\mathcal{F}_{q+1}^l$ is a set composed of the values of random variables from previous iterations and satisfies $\mathcal{F}_q^l \subset \mathcal{F}_{q+1}^l$. Therefore, we define $\mathcal{F}_q^l$ as an increasing sequence of $\sigma$-algebras \cite{measure-theory}
\begin{equation}\label{sigma algebra}
\mathcal{F}_q^l=\left\{t_1^1, t_2^1, \ldots, t_{m_1}^1, t_1^2, t_2^2, \ldots, t_{m_2}^2, t_1^l, \ldots, i_{q-1}^l\right\}
\end{equation}

Since $\mathbf{X}_q^l$ is calculated based on $t_1^1, t_2^1, \ldots, t_q^l$, and is measurable on $\mathcal{F}_{q+1}^l$ (where $\mathcal{F}_{q+1}^l$ is the measurable space and its functions are measurable sets), and $t_{q+1}^l$ is independent of $\mathcal{F}_{q+1}^l$, we have
\begin{equation}\label{expectation gradient}
\begin{aligned}
&\mathbb{E}\left[\xi_{q+1}^l \mid \mathcal{F}_{q+1}^l\right]  \\ &=\mathbb{E}_{t_{q+1}^l}\left[\xi_{q+1}^l\right] \\
& =\mathbb{E}_{t_{q+1}^l}\left[\operatorname{grad} f_{t_{q+1}^l}\left(\mathbf{X}_q^l\right)-\mathcal{T}_{\widetilde{\eta}_q^l}\left(\operatorname{grad} f_{t_{q+1}^l}\left(\widetilde{\mathbf{X}}^{l-1}\right)\right) \right. \\
&\left. \ \ + \mathcal{T}_{\widetilde{\eta}_t^l}\left(\operatorname{grad} f\left(\widetilde{\mathbf{X}}^{l-1}\right)\right)\right]
\end{aligned}
\end{equation}

According to the linearity property of $\mathcal{T}_{\widetilde{\eta}_q^l}$, we have
\begin{equation}\label{expectation vector transport}
\begin{aligned}
&\mathbb{E}_{t_{q+1}^l}\left[\mathcal{T}_{\tilde{\eta}_q^l}\left(\operatorname{grad} f_{t_{q+1}^l}\left(\widetilde{\mathbf{X}}^{l-1}\right)\right)\right]\\
& =\mathcal{T}_{\tilde{\boldsymbol{\eta}}_q^l}\left(\mathbb{E}_{t_{q+1}^l}\left[\operatorname{grad} f_{t_{q+1}^l}\left(\widetilde{\mathbf{X}}^{l-1}\right)\right]\right) \\
& =\mathcal{T}_{\tilde{\boldsymbol{\eta}}_q^l}\left(\operatorname{grad} f\left(\widetilde{\mathbf{X}}^{l-1}\right)\right)
\end{aligned}
\end{equation}

Substituting \eqref{expectation vector transport} into \eqref{expectation gradient}, we get
\begin{equation}\label{expectation gradient simple}
\mathbb{E}\left[\xi_{q+1}^l \mid \mathcal{F}_{q+1}^l\right]=\operatorname{grad} f\left(\mathbf{X}_q^l\right)
\end{equation}

Therefore, we have
\begin{equation}\label{expectation inner product}
\mathbb{E}\left[\left\langle\xi_{q+1}^l, \operatorname{grad} f\left(\mathbf{X}_q^l\right)\right\rangle_{\mathbf{x}_q^l} \mid \mathcal{F}_{q+1}^l\right]=\left\|\operatorname{grad} f\left(\mathbf{X}_q^l\right)\right\|_{\mathbf{X}_q^l}^2
\end{equation}

Then, \eqref{expectation inner product} leads to the following inequality
\begin{equation}\label{fX decrease bound}
\begin{aligned}
&\mathbb{E}\left[f\left(\mathbf{X}_{q+1}^l\right)-f\left(\mathbf{X}_q^l\right) \mid \mathcal{F}_{q+1}^l\right] \leq \\
&-\alpha_q^l\left\|\operatorname{grad} f\left(\mathbf{X}_q^l\right)\right\|_{\mathbf{X}_q^l}^2+\left(\alpha_q^l\right)^2 k_1 \leq\left(\alpha_q^l\right)^2 k_1
\end{aligned}
\end{equation}

From \eqref{fX decrease bound}, we have
\begin{equation}\label{fX bound}
\mathbb{E}\left[f\left(\mathbf{X}_{q+1}^l\right) \mid \mathcal{F}_{q+1}^l\right] \leq \mathbb{E}\left[f\left(\mathbf{X}_q^l\right) \mid \mathcal{F}_{q+1}^l\right]+\left(\alpha_q^l\right)^2 k_1
\end{equation}

Because $\mathcal{F}_{q+1}^l$ contains the information required to compute $f\left(\mathbf{X}_q^l\right)$, we have $\mathbb{E}\left[f\left(\mathbf{X}_q^l\right) \mid \mathcal{F}_{q+1}^l\right]=f\left(\mathbf{X}_q^l\right)$, and furthermore we have
\begin{equation}\label{further fX bound}
\mathbb{E}\left[f\left(\mathbf{X}_{q+1}^l\right) \mid \mathcal{F}_{q+1}^l\right] \leq f\left(\mathbf{X}_q^l\right)+\left(\alpha_q^l\right)^2 k_1
\end{equation}

The sequence  $\left\{\mathbf{X}_0^1, \mathbf{X}_1^1, \ldots, \mathbf{X}_{m_1}^1\left(=\mathbf{X}_0^2\right), \mathbf{X}_1^2, \mathbf{X}_2^2, \ldots, \right.$
$\left. \mathbf{X}_{m_2}^2, \ldots, \mathbf{X}_q^l, \ldots\right\}$ is relabeled as $\left\{\mathbf{X}_1, \mathbf{X}_2, \ldots, \mathbf{X}_{q_1}, \ldots\right\}$, and similarly relabel $\left\{\alpha_q^l\right\}$ and $\mathcal{F}_{q+1}^l$. According to \eqref{further fX bound}, we have
\begin{equation}\label{supermartingale}
\begin{aligned}
&\mathbb{E}\left[f\left(\mathbf{X}_{q_1+1}\right)+\sum_{l_1=q_1+1}^{\infty}\left(\alpha_{l_1}\right)^2 k_1 \mid \mathcal{F}_{q_1+1}\right] \\
&\leq f\left(\mathbf{X}_{q_1}\right)+\sum_{l_2=q_1}^{\infty}\left(\alpha_{l_2}\right)^2 k_1
\end{aligned}
\end{equation}

Based on equation \eqref{stepsize condition 1}, it is straightforward to obtain $\mathbb{E}\left[\left|f\left(\mathbf{X}_{q_1}\right)+\sum_{l_2=q_1}^{\infty}\left(\alpha_{l_2}\right)^2 k_1\right|\right]<\infty$. According to \cite{stochastic-processes}, \eqref{supermartingale} indicates that the stochastic process $\left\{f\left(\mathbf{X}_{q_1}\right)+\sum_{l_2=q_1}^{\infty}\left(\alpha_{l_2}\right)^2 k_1\right\}$ is a supermartingale. The problem considered in this paper satisfies $f\left(\mathbf{X}_{q_1}\right) \geq 0$, and in this case, $\left\{f\left(\mathbf{X}_{q_1}\right)+\sum_{l_2=q_1}^{\infty}\left(\alpha_{l_2}\right)^2 k_1\right\}$ is a non-negative supermartingale.

According to the martingale convergence theorem in [37], there exists a finite limit with probability 1, denoted as  $\lim _{q_1 \rightarrow \infty} f\left(\mathbf{X}_{q_1}\right)+\sum_{l_2=q_1}^{\infty}\left(\alpha_{l_2}\right)^2 k_1$, implying that the limit  $\lim _{q_1 \rightarrow \infty} f\left(\mathbf{X}_{q_1}\right)$ exists, $\left\{f\left(\mathbf{X}_q^l\right)\right\}$ converges.

 {\it Remark:} Up to this point, we have only proved that $\left\{f\left(\mathbf{X}_q^l\right)\right\}$ converges, i.e. the algorithm converges, but we have not yet specified the situation regarding the limit points, that is, the goodness of the solution. Next, we will prove that the algorithm converges to the stationary point.

Before proving the stationary point, we introduce a theorem:
\begin{theorem}\label{theorem-quasi-martingale}
 Let $\{X_n\}_{n \in \mathbb{N}}$ be a non-negative stochastic process with bounded positive variations, meaning it satisfies: $\sum_{n=0}^{\infty} \mathbb{E}\left[\mathbb{E}\left[X_{n+1}-X_n \mid \mathcal{F}_n\right]^{+}\right]<\infty$, where $X^{+}$ denotes the positive part of a random variable $X$, defined as $\max \{X, 0\}$, and $\mathcal{F}_n$ represents the increasing sequence of $\sigma$-algebras generated before time $n$. In this case, the stochastic process $\{X_n\}_{n \in \mathbb{N}}$ is a quasi martingale, i.e.,\cite{quasi-martingales}
\begin{equation}\label{quasi-martingale}
\sum_{n=0}^{\infty}\left|\mathbb{E}\left[X_{n+1}-X_n \mid \mathcal{F}_n\right]\right|<\infty
 \end{equation}
and $ X_n$ convergens a.s.
\end{theorem}

We accumulate inequality \eqref{fX decrease bound} from the current $q, l$ to $l \rightarrow \infty$ as follows
\begin{equation}\label{step grad inequation}
\begin{aligned}
&\sum_{l, q} \alpha_q^l\left\|\operatorname{grad} f\left(\mathbf{X}_q^l\right)\right\|_{\mathbf{X}_q^l}^2 \leq \\
&-\sum_{l, q} \mathbb{E}\left[f\left(\mathbf{X}_{q+1}^l\right)-f\left(\mathbf{X}_q^l\right) \mid \mathcal{F}_{q+1}^l\right]+\sum_{l, q}\left(\alpha_q^l\right)^2 k_1
\end{aligned}
\end{equation}

Here, by proving that the right-hand side of the inequality is bounded, we obtain the conclusion that the left-hand side of the inequality converges.

From \eqref{fX decrease bound}, we have
\begin{equation}\label{step grad bound}
\begin{aligned}
\sum_{l, q} \mathbb{E}\left[\mathbb{E}\left[f\left(\mathbf{X}_{q+1}^l\right)-f\left(\mathbf{X}_q^l\right) \mid \mathcal{F}_{q+1}^l\right]^{+}\right] &\leq \sum_{l, q}\left(\alpha_q^l\right)^2 k_1 \\
& < \infty
\end{aligned}
\end{equation}

According to {\it Theorem \ref{theorem-quasi-martingale}} we have
\begin{equation}\label{absolute value bound}
\begin{aligned}
&\left|-\sum_{l, q} \mathbb{E}\left[f\left(\mathbf{X}_{q+1}^l\right)-f\left(\mathbf{X}_q^l\right) \mid \mathcal{F}_{q+1}^l\right]\right|\\
 &\leq \sum_{l, q}\left|\mathbb{E}\left[f\left(\mathbf{X}_{q+1}^l\right)-f\left(\mathbf{X}_q^l\right) \mid \mathcal{F}_{q+1}^l\right]\right|\\
 &<\infty
\end{aligned}
\end{equation}

Therefore, we have
\begin{equation}\label{step grad convergence}
\sum_{l, q} \alpha_q^l\left\|\operatorname{grad} f\left(\mathbf{X}_q^l\right)\right\|_{\mathbf{X}_q^l}^2<\infty
\end{equation}
and $\sum_{l, q} \alpha_q^l\left\|\operatorname{grad} f\left(\mathbf{X}_q^l\right)\right\|_{\mathbf{X}_q^l}^2$ converges a.s. Additionally, $\lim _{l \rightarrow \infty} \alpha_q^l\left\|\operatorname{grad} f\left(\mathbf{X}_q^l\right)\right\|_{\mathbf{x}_q^l}^2=0$. Since the step size $\alpha_q^l$ decays, it does not imply the convergence of $\|\operatorname{grad} f\left(\mathbf{X}_q^l\right)\|_{\mathbf{X}_q^l}^2$. Next, we prove the convergence of $\|\operatorname{grad} f\left(\mathbf{X}_q^l\right)\|_{\mathbf{X}_q^l}^2$.

Here, following reference \cite{R-SGD}, we analyze the non-negative stochastic process $p_q^l=\|\operatorname{grad} f\left(\mathbf{X}_q^l\right)\|_{\mathbf{X}_q^l}^2$. Since $\|\operatorname{grad} f\left(\mathbf{X}_q^l\right)\|_{\mathbf{X}_q^l}^2$ is bounded, there exists an upper bound $k_2$ on the maximum eigenvalue of the Hessian matrix of $\|\operatorname{grad} f\left(\mathbf{X}_q^l\right)\|_{\mathbf{X}_q^l}^2$ along the curve connecting $\mathbf{X}_q^l$ and $\mathbf{X}_{q+1}^l$ on the compact set $K$. Using Taylor expansion
\begin{equation}\label{Taylor expansion second order}
\begin{aligned}
p_{q+1}^l-p_q^l &\leq-2 \alpha_q^l\left\langle\operatorname{grad} f\left(\mathbf{X}_q^l\right), \operatorname{Hess} f\left(\mathbf{X}_q^l\right)\left[\xi_{q+1}^l\right]\right\rangle_{\mathbf{X}_q^l}\\
&+ \left(\alpha_q^l\right)^2\left\|\xi_{q+1}^l\right\|_{\mathbf{X}_q^l}^2 k_2
\end{aligned}
\end{equation}

Let $-k_3$ be the lower bound of the minimum eigenvalue of the Hessian matrix of $f$ on $K$. Therefore, we have
\begin{equation}\label{gradient decrease bound}
\mathbb{E}\left[p_{q+1}^l-p_q^l \mid \mathcal{F}_{q+1}^l\right] \leq 2 \alpha_q^l\left\|\operatorname{grad} f\left(\mathbf{X}_q^l\right)\right\|_{\mathbf{X}_q^l}^2 k_3+\left(\alpha_q^l\right)^2 C^{\prime} k_2
\end{equation}

Previously, we have proven that
\begin{equation}\label{step gradient convergence 1}
\sum_{l, q} \alpha_q^l\left\|\operatorname{grad} f\left(\mathbf{X}_q^l\right)\right\|_{\mathbf{X}_q^l}^2<\infty
\end{equation}
\begin{equation}\label{step gradient convergence 2}
\lim _{l \rightarrow \infty} \alpha_q^l\left\|\operatorname{grad} f\left(\mathbf{X}_q^l\right)\right\|_{\mathbf{X}_q^l}^2=0
\end{equation}

Therefore, we can get
\begin{equation}\label{gradient convergence}
\sum_{l, q} \mathbb{E}\left[\mathbb{E}\left[p_{q+1}^l-p_q^l \mid \mathcal{F}_{q+1}^l\right]^{+}\right]<\infty
\end{equation}
and $\{p_q^l\}$  is a quasi martingale. According to {\it Theorem \ref{theorem-quasi-martingale}}, $\{p_q^l\}$ converges. According to equations \eqref{stepsize condition 1} and \eqref{stepsize condition 2}, $\lim _{l \rightarrow \infty} \alpha_q^l \rightarrow 0$ but not equal to 0. Therefore, $\{p_q^l\}$ converges to 0, indicating that the algorithm converges to a stationary point. This concludes the proof. $\hfill\blacksquare$

%
%
%





\begin{thebibliography}{99}

\bibitem{wireless-security-survey}
Y. Zou, J. Zhu, X. Wang and L. Hanzo, ``A Survey on Wireless Security: Technical Challenges, Recent Advances, and Future Trends," \emph{Proceedings of the IEEE}, vol. 104, no. 9, pp. 1727-1765, Sept. 2016.
\bibitem{Steganography}
J. Fridrich, \emph{Steganography in Digital Media: Principles, Algorithms, and Applications}, 1st ed. New York, NY, USA: Cambridge University Press, 2009.
\bibitem{covert-fundamental-limits}
B. A. Bash, D. Goeckel, D. Towsley and S. Guha, ``Hiding information in noise: fundamental limits of covert wireless communication," \emph{IEEE Communications Magazine}, vol. 53, no. 12, pp. 26-31, Dec. 2015.
\bibitem{square-root-law}
B. A. Bash, D. Goeckel and D. Towsley, ``Limits of Reliable Communication with Low Probability of Detection on AWGN Channels," \emph{IEEE Journal on Selected Areas in Communications}, vol. 31, no. 9, pp. 1921-1930, Sept. 2013.
\bibitem{SRL-BSC}
P. H. Che, M. Bakshi and S. Jaggi, ``Reliable deniable communication: Hiding messages in noise," in \emph{IEEE International Symposium on Information Theory}, pp. 2945-2949, July 2013.
\bibitem{DMC-SRL}
M. R. Bloch, ``Covert Communication Over Noisy Channels: A Resolvability Perspective," \emph{IEEE Transactions on Information Theory}, vol. 62, no. 5, pp. 2334-2354, May 2016.
\bibitem{DMC-SRL-Zheng}
L. Wang, G. W. Wornell and L. Zheng, ``Fundamental Limits of Communication With Low Probability of Detection," \emph{IEEE Transactions on Information Theory}, vol. 62, no. 6, pp. 3493-3503, June 2016.
\bibitem{MAC-SRL}
K. S. K. Arumugam and M. R. Bloch, ``Keyless covert communication over Multiple-Access Channels," in \emph{IEEE International Symposium on Information Theory}, pp. 2229-2233, July 2016.
\bibitem{Broadcost-SRL}
V. Y. F. Tan and S. -H. Lee, ``Time-Division is Optimal for Covert Communication Over Some Broadcast Channels," \emph{IEEE Transactions on Information Forensics and Security}, vol. 14, no. 5, pp. 1377-1389, May 2019.
\bibitem{MIMO-AWGN-SRL}
A. Abdelaziz and C. E. Koksal, ``Fundamental limits of covert communication over MIMO AWGN channel," in \emph{IEEE Conference on Communications and Network Security}, pp. 1-9, Oct. 2017.
\bibitem{transmission-time}
B. A. Bash, D. Goeckel and D. Towsley, ``Covert Communication Gains From Adversary's Ignorance of Transmission Time," \emph{IEEE Transactions on Wireless Communications}, vol. 15, no. 12, pp. 8394-8405, Dec. 2016.
\bibitem{channel-noise-power}
S. Lee and R. J. Baxley, ``Achieving positive rate with undetectable communication over AWGN and Rayleigh channels," in \emph{IEEE International Conference on Communications}, pp. 780-785, June 2014.
\bibitem{environment_interference}
R. Soltani, D. Goeckel, D. Towsley, B. A. Bash and S. Guha, ``Covert Wireless Communication With Artificial Noise Generation," \emph{IEEE Transactions on Wireless  Communications}, vol. 17, no. 11, pp. 7252-7267, Nov. 2018.	
\bibitem{continuous-time}
K. Li, T. V. Sobers, D. Towsley and D. Goeckel, ``Covert Communication in Continuous-Time Systems in the Presence of a Jammer," \emph{IEEE Transactions on Wireless Communications}, vol. 21, no. 7, pp. 4883-4897, July 2022.
\bibitem{noise-uncertainty}
B. He, S. Yan, X. Zhou and V. K. N. Lau, ``On Covert Communication With Noise Uncertainty," \emph{IEEE Communications Letters}, vol. 21, no. 4, pp. 941-944, April 2017.
\bibitem{poisson-interference}
B. He, S. Yan, X. Zhou and H. Jafarkhani, ``Covert Wireless Communication With a Poisson Field of Interferers," \emph{IEEE Transactions on Wireless Communications}, vol. 17, no. 9, pp. 6005-6017, Sept. 2018.
\bibitem{MIMO-Poisson}
T. -X. Zheng, H. -M. Wang, D. W. K. Ng and J. Yuan, ``Multi-Antenna Covert Communications in Random Wireless Networks," \emph{IEEE Transactions on Wireless Communications}, vol. 18, no. 3, pp. 1974-1987, March 2019.
\bibitem{non-cooperative}
T. -X. Zheng, Z. Yang, C. Wang, Z. Li, J. Yuan and X. Guan, ``Wireless Covert Communications Aided by Distributed Cooperative Jamming Over Slow Fading Channels," \emph{IEEE Transactions on Wireless Communications}, vol. 20, no. 11, pp. 7026-7039, Nov. 2021.	
\bibitem{cognitive-jammer}
W. Xiong, Y. Yao, X. Fu and S. Li, ``Covert Communication With Cognitive Jammer," \emph{IEEE Wireless Communications Letters}, vol. 9, no. 10, pp. 1753-1757, Oct. 2020.
\bibitem{multi-antenna-jammer}
O. Shmuel, A. Cohen and O. Gurewitz, ``Multi-Antenna Jamming in Covert Communication," \emph{IEEE Transactions on Communications}, vol. 69, no. 7, pp. 4644-4658, July 2021.
\bibitem{full-duplex-jammer}
K. Shahzad, X. Zhou, S. Yan, J. Hu, F. Shu and J. Li, ``Achieving Covert Wireless Communications Using a Full-Duplex Receiver," \emph{IEEE Transactions on Wireless Communications}, vol. 17, no. 12, pp. 8517-8530, Dec. 2018.
\bibitem{secure-MIMO}
J. Zhu, W. Xu and N. Wang, ``Secure Massive MIMO Systems With Limited RF Chains," \emph{IEEE Transactions on Vehicular Technology}, vol. 66, no. 6, pp. 5455-5460, June 2017.
\bibitem{AN-cooperative-relay}
D. Goeckel, S. Vasudevan, D. Towsley, S. Adams, Z. Ding and K. Leung, ``Artificial Noise Generation from Cooperative Relays for Everlasting Secrecy in Two-Hop Wireless Networks," \emph{IEEE Journal on Selected Areas in Communications}, vol. 29, no. 10, pp. 2067-2076, December 2011.
\bibitem{fading-channel-uncertain}
K. Shahzad, X. Zhou and S. Yan, ``Covert Communication in Fading Channels under Channel Uncertainty," in \emph{IEEE 85th Vehicular Technology Conference}, Sydney, NSW, Australia, pp. 1-5, 2017.
\bibitem{test-hypotheses}
E. L. Lehmann and J. P. Romano, \emph{Testing Statisitical Hypotheses.} New York, NY, USA: Springer, 2005.
\bibitem{information-theory}
Cover T M. \emph{Elements of information theory}. John Wiley \& Sons, 1999.
\bibitem{robust-beamforming}
S. Ma et al., ``Robust Beamforming Design for Covert Communications," \emph{IEEE Transactions on Information Forensics and Security}, vol. 16, pp. 3026-3038, 2021.
\bibitem{online-learning}
L. Bottou, ``Online learning and neural networks" in \emph{Online Algorithms and Stochastic Approximations}, U.K., Cambridge: Cambridge Univ. Press, 1998.
\bibitem{complex-stiefel}
H. Sato, ``Riemannian conjugate gradient method for complex singular value decomposition problem," \emph{53rd IEEE Conference on Decision and Control}, Los Angeles, CA, USA, 2014, pp. 5849-5854.
\bibitem{R-SVRG}
Sato H, Kasai H, Mishra B, ``Riemannian Stochastic Variance Reduced Gradient Algorithm with Retraction and Vector Transport". \emph{SIAM Journal on Optimization}, vol. 29, no. 2, pp. 1444-1472, 2019.
\bibitem{SVRG}
Johnson R, Zhang T . ``Accelerating stochastic gradient descent using predictive variance reduction". \emph{News in Physiological Sciences}, vol. 1, no. 3, pp. 315-323, 2013.
\bibitem{manifold-optimization-absil}
P.-A. Absil, R. Mahony and R. Sepulchre, \emph{Optimization Algorithms on Matrix Manifolds}, Princeton, NJ, USA:Princeton Univ. Press, 2009.
\bibitem{topology}
James R M. \emph{Topology}. Prentic Hall of India Private Limited, New delhi, 2000.
\bibitem{stiefel-manifold-tsp}
T. Kaneko, S. Fiori and T. Tanaka, ``Empirical Arithmetic Averaging Over the Compact Stiefel Manifold," \emph{IEEE Transactions on Signal Processing}, vol. 61, no. 4, pp. 883-894, Feb.15, 2013.
\bibitem{measure-theory}
Halmos P R. \emph{Measure theory}. Springer, 2013.
\bibitem{stochastic-processes}
Ross S M. \emph{Stochastic processes}. John Wiley \& Sons, 1995.
\bibitem{super-martingale}
Durrett R. \emph{Probability: theory and examples}. Cambridge university press, 2019.
\bibitem{quasi-martingales}
Fisk D L. ``Quasi-martingales," \emph{Transactions of the American Mathematical Society}, vol. 120, no. 3, pp. 369-389, 1965.
\bibitem{R-SGD}
S. Bonnabel, ``Stochastic Gradient Descent on Riemannian Manifolds," \emph{IEEE Transactions on Automatic Control}, vol. 58, no. 9, pp. 2217-2229, Sept. 2013.




\end{thebibliography}
\end{document}